**A new radial basis function collocation method based on the quasi-uniform nodes for 2D fractional wave equation**

**Yiran Xu[1], Jingye Li[1,*], Guofei Pang[2], Zhikai Wang[1], Xiaohong Chen[1], Benfeng Wang[3]**

1. State Key Laboratory of Petroleum Resources and Prospecting, National Engineering Laboratory for Offshore Oil Exploration, China University of Petroleum-Beijing, Changping 102249, Beijing, China.

2. Division of Applied Mathematics, Brown University, Providence, RI 02912

3. State Key Laboratory of Marine Geology, School of Ocean and Earth Science, Institute for Advanced Study, Tongji University, Shanghai, 200092, China.

**Corresponding Author: Jingye Li**

**Email Address: ljy3605@sina.com**

**Abstract**

We mainly concerned with a decoupled fractional Laplacian wave equation in this paper. A new time-space domain radial basis function (RBF) collocation method is introduced to solve the fractional wave equation, which describes seismic wave propagation in attenuation media. The directional fractional Laplacian is adopted to cope with the fractional Laplacian of RBFs. In order to increase the computational efficiency, we introduced the quasi-uniform nodes configuration scheme, which is suitable for mesh-free discretization of wave equations. The comparison between the new method and the commonly-used pseudo-spectral method are implemented on square homogeneous models with different model size. The CPU time and relative errors of different methods show that the quasi-uniform configuration scheme provides better performance and the calculation efficiency advantage is significantly prominent as the computation domain increases. The relative errors indicate that the RBF collocation method with quasi-uniform configuration could improve the computational efficiency effectively and provide satisfactory accuracy. This advantage was especially highlighted in complex models, where the new approach achieved the same accuracy with only a half number of points. The implementation on the 2D complex model further demonstrated the accuracy, efficiency, and flexibility of the proposed new method.

**Keyword**: radial basis function; quasi-uniform configuration; fractional Laplacian; numerical modelling.

## 1. Introduction

Fractional differential equations are widely used to describe the propagation of seismic waves in viscoelastic media [1-4]. The fractional derivative constitutive model can not only describe the nonlinear long-term effect of deformation recovery of viscoelastic materials well, but also describe its unique stress relaxation and creep phenomena [5]. Among numerous numerical methods for solving the fractional derivative equations, there are two main types of mesh-based local approximation methods, such as the finite-difference method (FDM) and spectral-element method (SEM). However, for the FDM, the local approximation scheme loses the merit of the sparse resultant owing to the non-local property of the space-fractional derivative. Moreover, the requirement of the regular grids lacks the geometric flexibility to adjust to nonplanar surface and interfaces. The SEM is a high order finite-element (FE) technique that combines the geometric flexibility of the FE with the high accuracy of the spectral method [6, 7]. Although the FE method can handle curved boundaries and arbitrarily shaped anomalies directly by using irregular grids [8], the irregular grids require a complex meshing process as well as more memory and computational cost than the standard FDMs.

Currently, increasing literatures focus on radial-basis-function (RBF) based approximation methods [9-12] to realize mesh-free discretization. Without lattice mapping or complex meshing process, the mesh-free methods form the mesh elements by scattered nodes, which are originally adopted in the weak-form schemes [13-15]. Recently, the so-called RBF-FD method [16, 17], which combines RBF with the finite difference, has been proved to be the ideal method to obtain local derivative approximation with a high accuracy. The RBF-FD method extended to solve elastic wave equations and is developed further to provide third-order accuracy not only for the smooth regions, but also for reflections and transmissions at arbitrarily curved material interfaces [18, 19]. An alternative approach based on minimizing the absolute error of the dispersion relation by least-squares (LS) optimization is proposed for 2D acoustic-wave modelling [20]. However, the computational cost of solving a stable matrix by LS cannot be ignored. Besides, the above-mentioned methods only solve the wave equation with integer spatial differential, rather than the fractional spatial Laplacian wave equation.

For the fractional partial derivative equations (PDEs), the mesh-free methods belonging to global scheme, have certain advantages in the numerical simulation by virtue of their high accuracy and smaller size of the resultant matrix equation [21]. The mesh-free nodal distribution is model adaptive with respect to irregular structures and model parameters which is achieved by placing scattered computational nodes [20]. Although the mesh-free discretization scheme avoids the burden of grid generation for the irregular interface, these global schemes are still computationally expensive as compared with the local FDMs. Therefore, the variable density quasi-uniform node layout scheme was introduced to the modelling algorithm [22] in this paper. For the same size of the calculation domain, the quasi-uniform nodes layout scheme can decrease the number of calculation points efficiently as comparing to the regular rectangular grid.

The time domain fractional Laplacian visco-acoustic wave equation [23] has been provided for modelling the attenuation effect. Therein, the fractional Laplacians are evaluated in the frequency-space domain using a Fourier pseudo-spectral implementation. Laterly, two FFT-based modelling schemes are proposed for the solution of fractional Laplacian visco-acoustic equation [24]. The first scheme adopts the weighted sum of two constant-order fractional Laplacian to approximate the spatial variable-order fractional Laplacian and the second scheme introduces the low-rank decomposition [25] to approximate the mixed-domain propagator. However, those modelling schemes are both based on the Fourier transform. The FFT-based methods require the global nodes to perform the Fourier transforms at every time step, which costs a heavy computational burden, especially in the 3D modelling. Furthermore, the Gibbs phenomenon of Fourier transform results in the difficulties to handle the artificial boundary. By contrast, the RBF collocation method solves the fractional wave equation in the time-space domain and can easily handle the discretization of high-dimensional and irregular computational domain, since the discretization only depends only on the node-to-node distance.

This paper is organized as follows. Starting from Zhu's decoupled fractional Laplacian wave equation with constant-$Q$ [23], we derive the semi-discretization scheme in the time-space domain by determining the coefficients of nodes with RBF approximation. After accuracy and efficiency analysis of our proposed algorithm, some numerical modelling examples are provided to demonstrate the flexibility of the new method. Finally, a conclusion is given in the ending.

## 2. Methodology

### 2.1 Governing wave equation

The near constant-$Q$ (NCQ) decoupled visco-acoustic wave equation [26] based on fractional Laplacian is expressed by

$$\frac{1}{c^2}\frac{\partial^2 u}{\partial^2 t} = \eta(-\nabla^2)^{\gamma+1}u + \tau\frac{\partial}{\partial t}(-\nabla^2)^{\gamma+\frac{1}{2}}u, \tag{2.1}$$

where the concentration $u(x,y,t)$ is the wavefield variable and $c = c_0\cos(\pi\gamma/2)$ is the wave velocity depending on the phase velocity $c_0$ at a reference frequency $\omega_0$. Accordingly, $\eta = -c_0^{2\gamma}\omega_0^{-2\gamma}\cos(\pi\gamma), \tau = -c_0^{2\gamma-1}\omega_0^{-2\gamma}\sin(\pi\gamma)$, the reference frequency is defined by $f_0 = \omega_0/(2\pi)$ and $\nabla^2$ represents the Laplacian operator. The first Laplacian item of the right-hand side affects the seismic wave phase, and the second item affects the amplitude attenuation. The parameter $\gamma = \arctan(1/Q)$ is dimensionless, and $Q$ describes loss mechanisms. It is notable that $0 < \gamma < 0.5$ for any positive value of the quality factor $Q$ (as shown in Fig. 1).

### 2.2 Review of RBF collocation method

Initially, the RBF collocation method was used to reconstruct hypersurfaces of multivariate functions [27]. Therein, the stress field $u(x,y,t)$ is approximated by the weighted sum of RBFs $\phi(r)$

$$u(x_i, y_i, t) \approx \sum_{j=1}^{M+N} \lambda_j(t)\phi(r_{ij}), \tag{2.2a}$$

$$r_{ij} = \sqrt{(x_i - x_j)^2 + (y_i - y_j)^2}, \tag{2.2b}$$

where $\{(x_j, y_j)\}$ is a group of source points located in the computational domain, $M$ and $N$ are numbers of points on the calculate domain and boundary, respectively. The RBF $\phi(\cdot)$ only depends on the distance (or the relative location) between the collocation point $(x_i, y_i)$ and the source points $(x_j, y_j)$, namely $r_{ij}$. Therefore the collocation of source points are not constrained by any mesh or element. $\lambda_j(t)$ are the time-dependent expansion coefficients to be evaluated.

Substitute the RBF approximation equations (2.2a) and (2.2b) into equation (2.1). Applying the Dirichlet boundary condition $u(\mathbf{x}, t) = 0, \mathbf{x} \in \Omega, t \in [0, T]$ to the homogeneous medium simulation and using the FD scheme for the temporal discretization, we can derive a set of semi-discretization equations

$$\frac{1}{c_i^2} \sum_{j=1}^{M+N} \frac{\lambda_j^{n+1} - 2\lambda_j^n + \lambda_j^{n-1}}{\Delta t^2} \phi_{ij} = \eta \sum_{j=1}^{M+N} \lambda_j^n (-\nabla^2)^{\gamma+1} \phi_{ij} + \tau \sum_{j=1}^{M+N} \frac{\lambda_j^{n+1} - \lambda_j^n}{\Delta t} (-\nabla^2)^{\gamma+1/2} \phi_{ij} + f_i^n, \quad (2.3a)$$

$$(x_i, y_i) \in \Omega, \quad i = 1, 2, \cdots, M, \quad (2.3b)$$

$$\sum_{j=1}^{M+N} \lambda_j \phi_{ij} = 0, \quad (x_i, y_i) \in \partial\Omega, \quad i = M+1, M+2, \cdots, M+N. \quad (2.3c)$$

The superscript of $\lambda_j^n$ indicates the RBF expansion coefficient for the $j$-th node at $(n-1)\Delta t$, where $\Delta t$ is the time step and $\phi_{ij}$ is defined by $\phi(r_{ij})$. $(-\nabla^2)^\beta \phi_{ij}$ represents $[(-\nabla^2)^\beta \phi](x_i, y_i)$, where $\beta$ can be $\gamma+1$ or $\gamma+1/2$. The first $M$ equations in (2.3a) correspond to the approximation of the governing equation and the last $N$ equations approximate the Dirichlet boundary condition. An initial wavefield $f_i^n = f(x_i, y_i, (n-1)\Delta t)$ is adopted for the external force term.

Rearranging the above approximating scheme, we have the matrix form

$$\begin{bmatrix} \mathbf{\Phi}_d - \tau \mathbf{C}^2 \mathbf{\Phi}_{\gamma+1/2} \\ \mathbf{\Phi}_b \end{bmatrix} \lambda^{n+1} = \begin{bmatrix} \eta \mathbf{C}^2 \mathbf{\Phi}_{\gamma+1} - \tau \Delta t \mathbf{C}^2 \mathbf{\Phi}_{\gamma+1/2} \\ \mathbf{\Phi}_b \end{bmatrix} \lambda^n + \begin{bmatrix} -\mathbf{\Phi}_d \\ 0 \end{bmatrix} \lambda^{n-1} + \begin{bmatrix} \Delta t \mathbf{C}^2 \mathbf{f}^n \\ 0 \end{bmatrix}. \quad (2.4)$$

Where $n = 2, 3, \ldots, L$, $T = L\Delta t$. $\mathbf{\Phi}_d$ is a matrix formed by the RBFs with domain-type collocation points and all the source points, i.e. $[\phi_{ij}]$ with $i = 1, 2, \cdots, M$; $j = 1, 2, \cdots, M+N$ Similarly, $\mathbf{\Phi}_b$ is given by $[\phi_{ij}]$ with $i = M+1, M+2, \cdots, M+N$; $j = 1, 2, \cdots, M+N$. $\mathbf{C}$ is

$M \times M$ diagonal matrix with the velocities at domain-type collocation points as the diagonal elements, namely $[c_i]$ with $i = 1, 2, \cdots, M$. Corresponding to the $n$-th time layer, $\boldsymbol{\lambda}^n = [\lambda_j^n]$ and $\mathbf{f}^n = [f_i^n]$ are $(M+N)$- and $M$- dimensional column vectors, respectively. $\boldsymbol{\Phi}_{\gamma+1}$ and $\boldsymbol{\Phi}_{\gamma+1/2}$ are the $M \times (M+N)$ matrices formed by the fractional derivatives of RBFs. The first expansion coefficient vector $\boldsymbol{\lambda}^1$ is obtained by solving the RBF interpolation problem

$$\begin{bmatrix} \Phi_d \\ \Phi_b \end{bmatrix} \boldsymbol{\lambda}^1 = u_0, \tag{2.5}$$

where $u_0 = u_0(x_i, y_i)$ represents the initial wavefield with $i = 1, 2, \cdots, M + N$. The second expansion coefficient vector $\boldsymbol{\lambda}^2$ is assumed as coincides with the first one. Through the iteration of the RBF expansion coefficients $\boldsymbol{\lambda}^n$ in (4) starting from $\boldsymbol{\lambda}^1, \boldsymbol{\lambda}^2$, we can derive the coefficients for any given time $t_n = (n-1)\Delta t$.

**2.3 Fractional Laplacian of RBF**

The fractional Laplacian and the fractional derivative are two related while different mathematical concepts. Both of them are defined through a singular convolution integral. The former is a positive definition via the Riesz potential as the standard Laplace operator, while the latter via the Riemann-Liouville integral is not [28, 29]. Note that the fractional Laplacian cannot be interpreted by the fractional derivative and the later in the sense of either Riemann-Liouville or Caputo. Both the fractional Laplacian and the fractional derivative have wide applications in many complicated engineering problems. However, the standard definition of the fractional Laplacian leads to a hyper-singular convolution integral and is also obscure about how to implement the boundary conditions. In this paper, we adopt the directional definition [30, 31] to approximate the fractional Laplacian of RBFs, which can be characterized by

$$(-\nabla^2)^{\alpha/2} u(x) = C_{a,d} \int_{\|\theta\|=1} D_\theta^\alpha u(x) d\theta, \qquad x, \theta \in \mathbb{R}^d, \tag{2.6}$$

where the scaling constant before the defined as,

$$C_{a,d} = \frac{\Gamma(\frac{1-a}{2})\Gamma(\frac{d+a}{2})}{2\pi^{\frac{1+d}{2}}}. \tag{2.7}$$

The fractional directional derivative is given by

$$D_\theta^\alpha(\cdot) = (\nabla \cdot \theta)^2 I_\theta^{2-\theta}(\cdot), \tag{2.8}$$

where $\nabla$ is the gradient operator and the fractional directional integral $I_\theta^\beta(\cdot)$ is defined as follows

$$I_\theta^\beta u(x) = \frac{1}{\Gamma(1-\beta)} \int_0^{+\infty} \varsigma^{-\beta} u(x - \varsigma\theta) d\varsigma, \quad \beta \in (0,1), \tag{2.9}$$

where $\Gamma(\cdot)$ is the Euler Gamma function. The directional definition is equivalent to the Riesz one and it is easy to approximate by using RBF collocation method. In 2D cases, we adopt the vector Grünwald scheme [32] to the directional definition

$$D_\theta^\beta \phi(r_{ij}) \approx h^{-\beta} \sum_{k=0}^{\left[\frac{d(x_i, y_i, \theta, \Omega)}{h}\right]} (-1)^k \binom{\beta}{k} \phi(\sqrt{(x_i - kh\cos\theta - x_j)^2 + (y_i - kh\sin\theta - y_j)^2}). \tag{2.10}$$

The symbol $[\alpha]$ means the closest integer to the fraction $\alpha$. $d(x_i, y_i, \theta, \Omega)$ is the distance from the collocation point $(x_i, y_i)$ to the boundary of computing domain along the direction $(\cos\theta, \sin\theta)$ (shown in Fig. 2), evaluating the integration of the above directional derivative with respect to $\theta$. The integral can be accurately approximated by using trapezoidal rule since $D_\theta^\beta(\bullet)$ is periodic for $\theta$. Thus, we have the quadrature formula

$$\begin{aligned}(-\nabla^2)^2 \phi(r_{ij}) &= C_{2\beta,2} \int_0^{2\pi} D_\theta^\beta \phi(r_{ij}) d\theta \\ &\approx \frac{2\pi C_{2\beta,2}}{N_t} \sum_{l=0}^{N_t-1} D_{\theta_l}^\beta \phi(r_{ij}), \quad (\theta = 2\pi l / N_t)\end{aligned} \tag{2.11}$$

where $N_t + 1$ denotes the used trapezoidal quadrature points are used. Letting the fractional order $\beta$ be $\gamma + 1$ and $\gamma + 1/2$, the index $i$ goes from 1 to $M$ and the index $j$ goes from 1 to $M + N$, we can compute all the elements of matrices $\Phi_{\gamma+1}$ and $\Phi_{\gamma+1/2}$ finally. As for the choice of RBF $\phi$, Table 1 shows various types of RBFs. Among various interpolation schemes, Multiquadric (MQ) and Thin plate spline (TPS) interpolations perform best [33]. In this paper, the MQ interpolation is used to interpolate the wavefield.

## 2.4 Variable density nodes layout scheme

Considering the globality of the fractional Laplacian, we use a variable density node layouts scheme [22] to improve numerical calculating efficiency. This advancing front type algorithm is suitable for mesh-free discretization of PDEs. This scheme starts by randomly scattering 'potential dot locations'(PDPs) relatively and densely along bottom edge of rectangle and then proceeds by the following steps:

> **Repeat** until lowest PDP is above the top edge of the domain
> 1. Find the lowest PDP and select it as a new dot location.
> 2. Mark out a circle centered at the new dot and with a radius matching the desired node separation distance at this location.
> 3. Keep the new dot, and remove all other PDPs within this circle.
> 4. Note the directions to the nearest remaining PDP on each side. On the circle periphery, place five new PDP equally spaced in angle within this sector.
>
> **End repeat**

The MATLAB code of the algorithm above can be found in the Appendix, proposed by Fornberg and Flyer [22]. Note that the radius above is RBG Parade-based computing. When it applied to the velocity model, the RGB Parade need to be converted to the velocity value. Fig. 3 shows the algorithm which applied in a 2D homogenous media, and the two different node configuration in a 2D homogeneous media is with size $0.5km \times 0.5km$. The square uniform nodes distribution scheme is with $dx = dz = 10m$, and it contains 2601 nodes and the quasi-uniform scheme contains 2080 nodes. In the quasi-uniform scheme, the basis node radius $r_{bs}$ is set depending on the velocity value of node location. When $v = 2000 m/s$, the basis node radius $r_{bs} \approx 10$.

## 3. Accuracy and efficiency analysis

We analyze the waveform error curves and CPU lapsed time of different methods to demonstrate the accuracy and efficiency of the new method. For simplicity, the 2D homogeneous model is shown to perform the analysis.

The velocity of the 2D homogeneous model is $2000 m/s$, with model size $1km \times 1km$. For the uniform grids, the grid spacing is set $10m$ to discretize the model. The source wavelet is a Ricker wavelet with a dominant frequency of 10Hz. We simulate the wave propagation using the RBF methods with the uniform grid (UG) and quasi-uniform grids (QUG) discretization, respectively. The result obtained by using the FFT-based pseudo-spectral method is considered as the reference solution. Fig. 4 shows some snapshots at $100ms$ and $200ms$, computed using different methods. Fig. 5 compares the waveform of trace at compares the waveform of trace at $x = 0.5km$. For the quasi-uniform nodes layout scheme, we interpolate the trace using Spline interpolation method at $x = 0.5km$. Fig. 6 provides a zoomed-in version of Fig.5 at $0.16 - 0.26km$ and $0.8 - 0.9km$. It appears that the waveform computed by RBF-QUG and RBF-UG almost coincides with the reference solutions.

To test the efficiency of RBF collocation method with quasi-uniform nodes layout scheme, we compare the calculation behavior of the two different node configuration schemes by using a homogeneous model. The space interval of the uniform grid is $dx = dz = 10m$, and the radius of quasi-uniform nodes is set as $1-2$ times to $r_{bs} = 15m$. Using different sizes models, we investigate the computational times and relative error which computed by $\varepsilon = \left|\phi_{RBF} - \phi_{ref}\right|/\phi_{ref}$, where $\phi_{RBF}$ is result computed by the RBF collocation methods and $\phi_{ref}$ is the solution by the pseudo-spectral method. The computations are all performed on the same computer (Dell with Intel(R) Xeon(R) CPU E5-2698, 2,20GHz, and 192 GB memory) using Matlab. Table 2 presents the CPU times and relative errors of RBF collocation method with two different node configuration schemes. Table 2 notes that the quasi-uniform configuration scheme provides better discrete performance and higher accuracy while the calculation efficiency is significantly improved as the computation domain increases.

According to equation (1), the decoupled fractional Laplacian wave equation describes very nearly constant-$Q$ attenuation and dispersion effects. We design a different $Q$ model to verify the flexibility of the proposed method. The velocity model is homogeneous as used in Fig. (3) and the source is located at the center of the model. Fig. 7 shows the snapshots at $200ms$ in the NCQ wave equation with $Q$ = 100, 30, 10 and 4. Three subpanels from left to right computed by RBF collocation method based on the quasi-uniform node distribution, RBF collocation method based on the uniform node distribution, and pseudo-spectral method, respectively. We can see decreased amplitude and delayed phase with decreasing $Q$ values.

## 4. Numerical examples

In the following section, we compare the proposed RBF collocation method with two different nodes layout schemes, and FFT based pseudo-spectral method to verify that the new method can improve the numerical accuracy under the sparse grids. First, we choose the partial Hess model and discrete it with different radius. The simulation results by different methods under different node layouts schemes verify the validity of the proposed method. Then, the applications to the 2D modified Marmousi model demonstrate the validity of the new method when compared to the pseudo-spectral method. For the 2D heterogeneous medium simulation, the exponential absorption boundary condition is used to attenuate the boundary reflections.

### 4.1 Partial SEG Hess Model

In this section, we perform the numerical simulations with the new method and the FFT based pseudo-spectral method on part of 2D SEG Hess model (Fig. 8a). Here, the 2D medium is extracted from the SEG Hess model. The model size is $3km \times 9km$. For the quasi-uniform scheme, we set three increasing radii to discrete the medium. And then we apply the pseudo-spectral method with the square grids, which have grid spacing to the quasi-uniform scheme approximately. The grids spacing of quasi-uniform layout is variable according to the velocity, so we chose an approximate value to verify the simulation results. Here the grid spacing is set as $7.5m$ and $30m$ for the pseudo-spectral method which is used for comparison with the RBF collocation method under the approximate radius.

Fig. 8b-8d present the quasi-uniform node layouts with increasing grid spacing. The grid spacing in Fig. 8b is $r_{bs} = 5 \sim 10m$ and the grids spacing in Fig. 8c and 8d are double and triple to $r_{bs}$. The total numbers of nodes in Fig. 8b-8d are 117412, 29494 and 13051 respectively. The total number of nodes in the regular grid discretization is 480000 with the grid spacing $7.5m$. When the grid spacing increases to $30m$, the total number of nodes decreases to 9000. The quality factor $Q$ is set as 100.

Fig. 9 displays some snapshots at $400ms$, $600ms$, and $800ms$. The Ricker wavelet source with dominant frequency of 16Hz is located at (4500, 3000). The temporal sample rate is $3ms$. Fig. 9a shows the reference solution, which is obtained by the pseudo-spectral method and the grid spacing of $7.5m$. Fig. 9b-9f show the snapshots calculated by RBF collocation method with regular square grids $dx = dz = 7.5m$, RBF collocation method with quasi-uniform nodes layout $r_{bs} = 5 \sim 10m$, RBF collocation method with quasi-uniform nodes layout $r = 2r_{bs}$, RBF collocation method with quasi-uniform nodes layout where $r = 3r_{bs}$, and FFT based pseudo-spectral method with sparse grid spacing where $dx = dz = 30m$, respectively. The results of the pseudo-spectral method with large grid spacing (Fig. 9f) suffer from serious spatial dispersion while the RBF collocation method with $r = 3r_{bs}$ yields better results. The different traces are shown in Fig. 10, Fig. 10a and 10c display the traces extracted from Fig. 9 at $400ms$. Fig. 10b and 10d show the difference of the various methods with reference result, which notes that our method outperforms the FFT based pseudo-spectral method with sparse grid spacing in accuracy. Meanwhile, the residuals show that the RBF collocation method with $r = 2r_{bs}$ can produce outstanding results of which need 1/16 nodes as the regular square grids of reference method.

**4.2 The 2D modified Marmousi model**

The last test is for the 2D modified marmousi model (Fig. 11a). The grid size is $280\times 680$ and the square grid is adopted for the pseudo-spectral method with the spatial interval 7.5m. The other relative parameters are displayed in the figure caption. Fig. 11b shows the non-uniform nodal distribution, and the node spacing ranges from 10 to 20m. The total number of nodes in Fig. 11b is 108446.

Fig. 12 displays the snapshots at $1s$ and $2s$ computed by RBF collocation method with quasi-uniform distribution and pseudo-spectral method based on the uniform node distribution. The time step is $2ms$. The Ricker wavelet source with 16Hz dominant frequency is used and receivers are located on $(0, 2500)$. Fig. 12a and 12b are obtained by the RBF collocation method while 12c and 12d are calculated by the FFT based pseudo-spectral method. Then, we compare these seismograms and their waveforms (as shown in Fig. 13). Fig. 13a and 13b show the seismic records acquired at the surface, the methods of these two records are corresponding to RBF collocation method and pseudo-spectral method. The magnified portion of seismograms are shown in Fig. 13c, where the red filled waveforms present the RBF collocation method and the black filled waveforms present the pseudo-spectral method. Fig. 13d and 13e show the comparison of waveform extracted from Fig. 13a and 13b (marked as the white line), and these details comparisonss indicate that the RBF collocation method is superior to the FFT based pseudo-spectral method. With the same accuracy requirements, the new method can save more computational resources.

## 5. Conclusions

We have developed a new time-space domain mesh-free solution for the fractional Laplacian visco-acoustic wave equation with constant quality factor $Q$ based on the radial basis function collocation method. The approximation of fractional Laplacian is determined by the directional derivative definition. Furthermore, we introduced the quasi-uniform node configuration scheme which significantly reduces the number of computational points. Numerical results based on the homogeneous/heterogeneous mediums verified that this scheme could increase computational efficiency effectively, especially when handling the complex models. The FFT based pseudo-spectral method with small grid spacing was used as reference solution to validate the accuracy of the proposed method. The results indicated that the RBF collocation method based on the quasi-uniform nodes layout obtains a higher precision than the pseudo-spectral method while using less computational points. Then, the numerical model in the 2D partial SEG Hess model and modified Marmousi model demonstrates the accuracy and efficiency of the new method. For the further research, the variable fractional-order need to be taken into account, because the quality factor of the real media is generally spatially dependent.

## 6. Acknowledgements

This work is financially supported by the National Natural Science Foundation of China (grant number: 41774129, 41774131), National Science and Technology Major Project (grant number: 2016ZX05024001-004), Science and Technology Project of CNPC (grant number: 2016A-3303).

**References**

[1] R. Schumer, M.M. Meerschaert, B. Baeumer, Fractional advection-dispersion equations for modeling transport at the Earth surface, Journal of Geophysical Research, 114 (2009).
[2] Q.Q. Li, H. Zhou, Q.C. Zhang, H.M. Chen, S.B. Sheng, Efficient reverse time migration based on fractional Laplacian viscoacoustic wave equation, Geophysical Journal International, 204 (2016) 488-504.
[3] W. Chen, S. Holm, Fractional Laplacian time-space models for linear and nonlinear lossy media exhibiting arbitrary frequency power-law dependency, Journal of the Acoustical Society of America, 115 (2004) 1424-1430.
[4] W. Chen, A new definition of the fractional Laplacian, Arxiv Cornell University Library, 69 (2002) 30–36.

[5] D. Yin, X. Duan, X. Zhou, Y. Li, Time-based fractional longitudinal–transverse strain model for viscoelastic solids, Mechanics of Time-Dependent Materials, 18 (2014) 329-337.
[6] A.T. Patera, A spectral element method for fluid dynamics: Laminar flow in a channel expansion, Journal of Computational Physics, 54 (1984) 468-488.
[7] C. Bernardi, M. Dauge, Y. Maday, Spectral Methods for Axisymmetric Domains, Gauthier-Villars, 1999.
[8] E. Chaljub, D. Komatitsch, J.P. Vilotte, Y. Capdeville, B. Valette, G. Festa, Spectral-element analysis in seismology, Advances in Geophysics, 48 (2008) 365-419.
[9] D. Stevens, H. Power, K.A. Cliffe, A meshless local RBF collocation method using integral operators for linear elasticity, International Journal of Mechanical Sciences, 88 (2014) 246-258.
[10] P. Farrell, H. Wendland, RBF MULTISCALE COLLOCATION FOR SECOND ORDER ELLIPTIC BOUNDARY VALUE PROBLEMS, SIAM Journal on Numerical Analysis, 51 (2013) 2403-2425.
[11] C. Fan, C. Chien, H. Chan, C. Chiu, The local RBF collocation method for solving the double-diffusive natural convection in fluid-saturated porous media, International Journal of Heat and Mass Transfer, 57 (2013) 500-503.
[12] Q.T.L. Gia, I.H. Sloan, H. Wendland, Multiscale RBF collocation for solving PDEs on spheres, Numerische Mathematik, 121 (2012) 99-125.
[13] G.H. Lee, H.J. Chung, C.K. Choi, Adaptive crack propagation analysis with the element-free Galerkin method, International Journal for Numerical Methods in Engineering, 56 (2003) 331-350.
[14] T.P. Fries, H.G. Matthies, Classification and Overview of Meshfree Methods, (2004).
[15] C. Wenterodt, O.V. Estorff, Dispersion analysis of the meshfree radial point interpolation method for the Helmholtz equation, International Journal for Numerical Methods in Engineering, 77 (2009) 1670-1689.
[16] V. Shankar, G.B. Wright, R.M. Kirby, A.L. Fogelson, A Radial Basis Function (RBF)-Finite Difference (FD) Method for Diffusion and Reaction---Diffusion Equations on Surfaces, J Sci Comput, 63 (2015) 745-768.
[17] V. Bayona, M. Moscoso, M. Carretero, M. Kindelan, RBF-FD formulas and convergence properties, Journal of Computational Physics, 229 (2010) 8281-8295.
[18] B. Martin, B. Fornberg, A. Stcyr, Seismic modeling with radial-basis-function-generated finite differences, Geophysics, 80 (2015).
[19] C. Zhang, R.J. Leveque, The immersed interface method for acoustic wave equations with discontinuous coefficients, Wave Motion, 25 (1997) 237-263.
[20] B. Li, Y. Liu, M.K. Sen, Z. Ren, Time-space-domain mesh-free finite difference based on least squares for 2D acoustic-wave modeling, GEOPHYSICS, 82 (2017) T143-T157.
[21] G.F. Pang, W. Chen, Z.J. Fu, Space-fractional advection-dispersion equations by the Kansa method, Journal of Computational Physics, 293 (2015) 280-296.
[22] B. Fornberg, N. Flyer, Fast generation of 2-D node distributions for mesh-free PDE discretizations, Comput Math Appl, 69 (2015) 531-544.
[23] T.Y. Zhu, J.M. Harris, Modeling acoustic wave propagation in heterogeneous attenuating media using decoupled fractional Laplacians, Geophysics, 79 (2014) T105-T116.
[24] H.M. Chen, H. Zhou, Q.Q. Li, Y.F. Wang, Two efficient modeling schemes for fractional Laplacian viscoacoustic wave equation, Geophysics, 81 (2016) T233-T249.

[25] S. Fomel, L. Ying, X. Song, Seismic wave extrapolation using lowrank symbol approximation, Geophysical Prospecting, 61 (2013) 526–536.
[26] T. Zhu, J.M. Harris, Modeling acoustic wave propagation in heterogeneous attenuating media using decoupled fractional Laplacians, GEOPHYSICS, 79 (2014) T105-T116.
[27] J. Fujiki, S. Akaho, Flexible Hypersurface Fitting with RBF Kernels, in, Springer Berlin Heidelberg, Berlin, Heidelberg, 2013, pp. 286-293.
[28] W. Chen, S. Holm, Fractional Laplacian, Levy stable distribution, and time-space models for linear and nonlinear frequency-dependent lossy media, arXiv: Mathematical Physics, (2002).
[29] F. Liu, M.M. Meerschaert, S. Momani, N.N. Leonenko, W. Chen, O.P. Agrawal, Fractional Differential Equations, International Journal of Differential Equations, 2010 (2010) 1-2.
[30] J.P. Roop, Computational aspects of FEM approximation of fractional advection dispersion equations on bounded domains in R2, J Comput Appl Math, 193 (2006) 243-268.
[31] G. Pang, W. Chen, K.Y. Sze, Gauss-Jacobi-type quadrature rules for fractional directional integrals, Comput Math Appl, 66 (2013) 597-607.
[32] M.M. Meerschaert, J. Mortensen, H.P. Scheffler, Vector Grünwald formula for fractional derivatives, Fractional Calculus & Applied Analysis, 7 (2004) 2004.
[33] R. Franke, Scattered Data Interpolation: Tests of Some Method, Mathematics of Computation, 38 (1982) 181-200.

**List of Figures**

Fig. 1. The backward distance $d(x,y,\theta)$ of node $(x,y)$ to the boundary of a rectangle in the direction $\theta$.

Fig. 2. Variation of the fractional order $\gamma$ with the increasing quality factor ($Q$), the formula $\gamma = \arctan(1/Q)/\pi$ indicates the $0 < \gamma < 0.5$ for any positive value of $Q$.

Fig. 3. Two different node configuration methods for same size area ($0.5km \times 0.5km$), the left one represents the uniform distribution scheme, it contains 2601 dots, and the right one represents the quasi-uniformly scheme, it contains 2080 dots. It appears that the quasi-uniformly scheme requires fewer points than the uniform configuration scheme. More importantly, this advantage will be more apparent with the increase of the computational domain.

Fig. 4. Snapshots shown by 'stem' function in Matlab. They were calculated in homogenous medium by two different node configuration schemes shown in Fig. 3. The left column shows the snapshots at 100ms and the right ones denote the results at 200ms. (a) and (b) computed by the RBF collocation method with quasi-uniformly node layout scheme, (c) and (d) computed by the RBF collocation method with square uniform node layout scheme. (e) and (f) computed by the pseudo-spectral method with square uniform node layout scheme.

Fig. 5. Trace comparision at $x = 0.5km$ in Fig. 4, (a) and (c) compare the waveforms that computed by RBF with uniform grids and pseudo-spectral method. (b) and (d) compare the waveforms that computed by RBF with quasi-uniform grids and pseudo-spectral method.

Fig. 6. The zoomed-in version of waveform Numerical solutions for three different node configuration schemes.

Fig. 7. Four wavefield parts corresponding to four $Q$ values: (a) 100, (b) 30, (c) 10 and (d) 4. The three subfigures from left to right computed by pseudo-spectral method, RBF collocation method based on the uniform node distribution and RBF collocation method based on the quasi-uniform node distribution, respectively. The snapshots are recorded at 200ms in homogeneous attenuating media.

Fig. 8. Complex model (a) and its quasi-uniform node configuration (b). (c), (d) and (e) show the snapshots at 400ms, 600ms, and 800ms. The black arrows indicate the high-velocity layer and its reflection, the appearance of the high-velocity layer reflected wave also proves the accuracy.

Fig. 9. The 2D partial SEG Hess model snapshots at $400ms$ (a1-f1), $600ms$ (a2-f2), $800ms$ (a3-f3). (a) The reference snapshots. (b) Snapshots computed by RBF collocation method with regular square uniform grids layout where $dx = dz = 7.5m$. (c) Snapshots computed by RBF collocation method with quasi-uniform grids layout where $r = r_{bs} = 5 \sim 10m$. (d) Snapshots computed by RBF collocation method with quasi-uniform grids layout where $r = 2r_{bs}$. (e) Snapshots computed by RBF collocation method with quasi-uniform grids layout where $r=3r_{bs}$. (f) Snapshots computed by pseudo-spectral method with regular square uniform grids layout where $dx=dz=30m$.

Fig. 10. Waveforms comparison among differenct methods for 2D partial SEG Hess model. (a) and (c) display the waveforms extracted from Fig. 9 at $400ms$, 10b and 10d show the difference of the various methods with reference result.

Fig. 11. The 2D Marmousi model and its quasi-uniform mesh-free nodal distribution. (a) The velocity model. (b) The quasi-uniform nodal distribution obtained by the fast-generation algorithm.[22]

Fig. 12. The 2D modified Marmousi snapshots at $1s$ and $2s$, (a, b) computed by RBF collocation method with quasi-uniform nodes distribution where $r=10 \sim 20m$ and (c, d) computed by pseudo-spectral method with regular square uniform nodes distribution where $dx=dz=7.5m$.

Fig. 13. (a-b) Seismograms from Fig. 10. (a) RBF method based on the quasi-uniform distribution of nodes and (b) pseudo-spectral method based on the uniform distribution of nodes. The subfigures (c) and (d) indicate the waveform comparison among different methods extracted from the (a) and (b) (white line indication). (c) Extracted from the left line and (d) extracted from the right line. (e) Local magnification of the white box in the seismograms (a,b) to highlight the local refraction waveforms. The red lines indicate the RBF method and the black lines indicate the pseudo-spectral method.

**List of Tables**

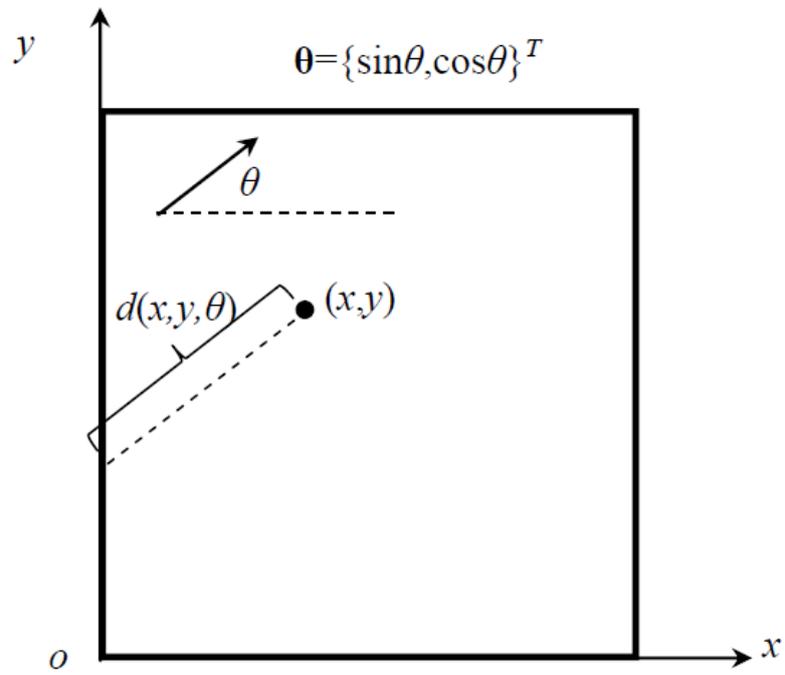

**Fig. 1.** The backward distance $d(x, y, \theta)$ of node $(x, y)$ to the boundary of a rectangle in the direction $\theta$.

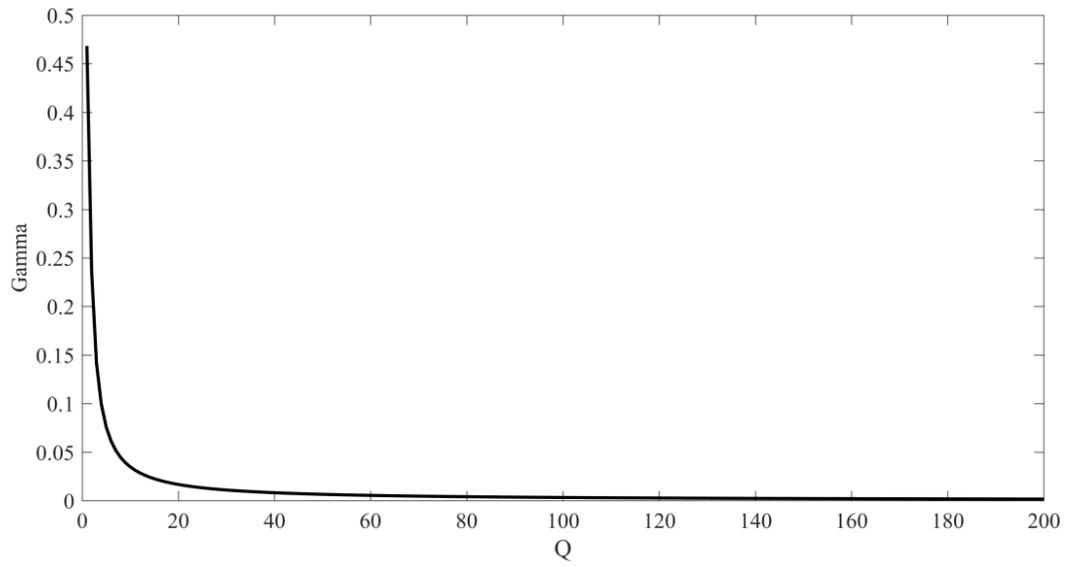

**Fig. 2.** Variation of the fractional order $\gamma$ with the increasing quality factor ($Q$), the formula $\gamma = \arctan(1/Q)/\pi$ indicates the $0 < \gamma < 0.5$ for any positive value of $Q$.

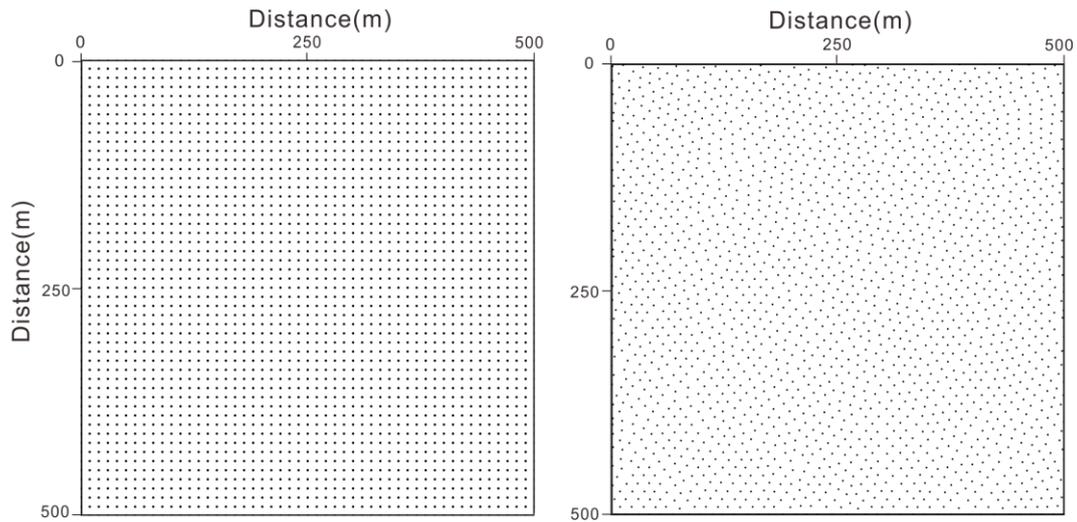

**Fig. 3.** Two different node configuration methods for same size area ($0.5km \times 0.5km$), the left one represents the uniform distribution scheme, it contains 2601 dots, and the right one represents the quasi-uniformly scheme, it contains 2080 dots. It appears that the quasi-uniformly scheme requires fewer points than the uniform configuration scheme. More importantly, this advantage will be more apparent with the increase of the computational domain.

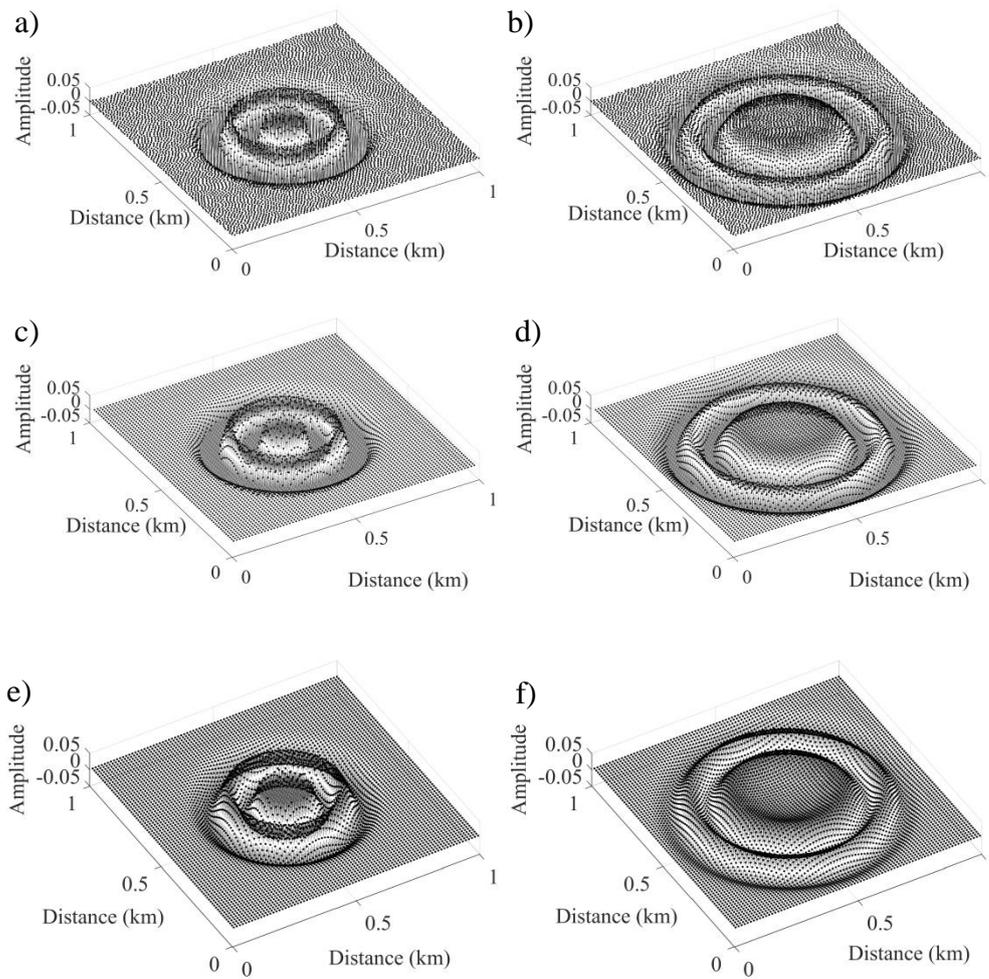

**Fig. 4.** Snapshots shown by 'stem' function in Matlab. They were calculated in homogenous medium by two different node configuration schemes shown in Fig. 3. The left column shows the snapshots at 100ms and the right ones denote the results at 200ms. (a) and (b) computed by the RBF collocation method with quasi-uniformly node layout scheme, (c) and (d) computed by the RBF collocation method with square uniform node layout scheme. (e) and (f) computed by the pseudo-spectral method with square uniform node layout scheme.

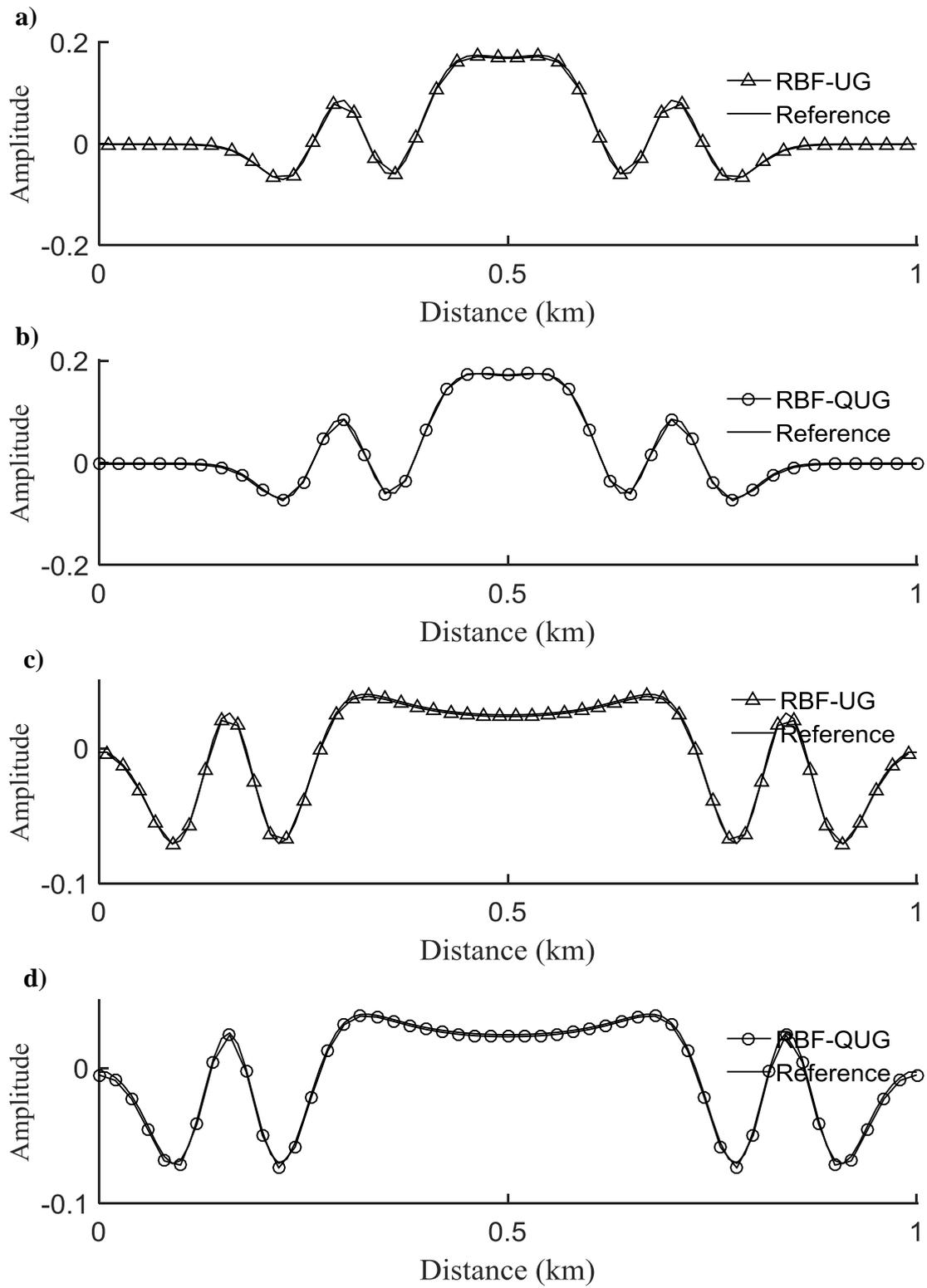

**Fig. 5.** Trace comparision at $x = 0.5km$ in Fig. 4, (a) and (c) compare the waveforms that computed by RBF with uniform grids and pseudo-spectral method. (b) and (d)

compare the waveforms that computed by RBF with quasi-uniform grids and pseudo-spectral method.

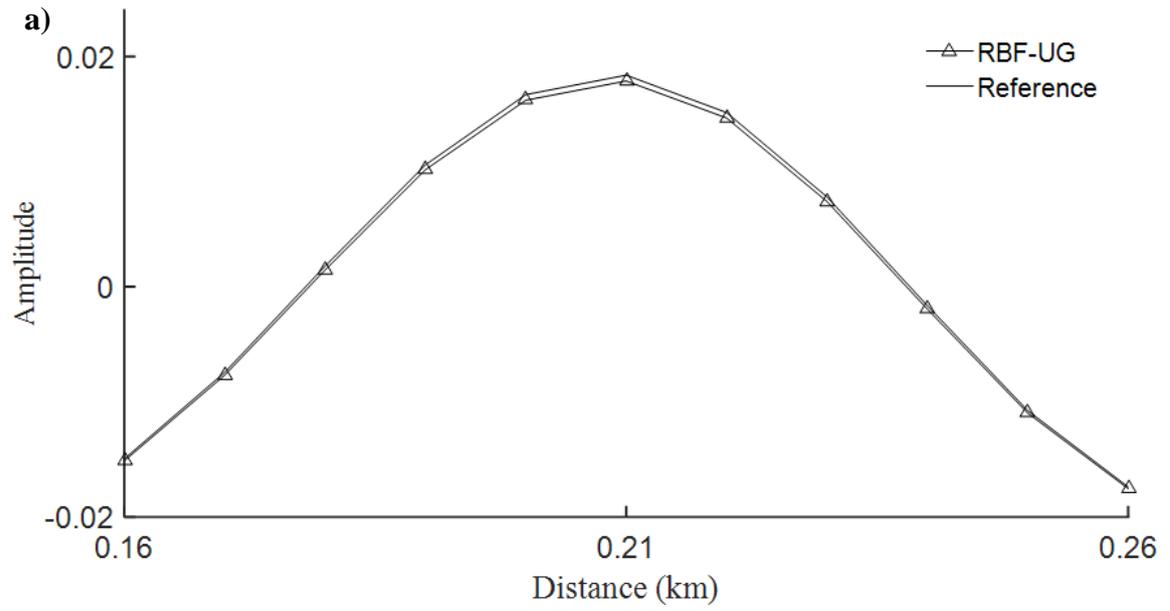
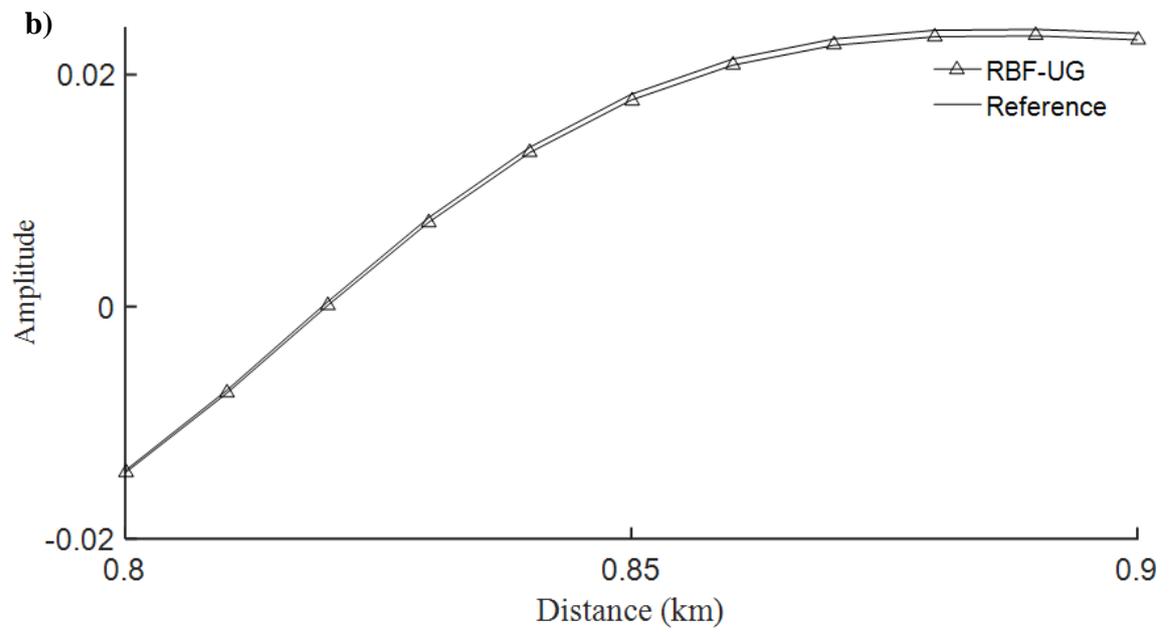

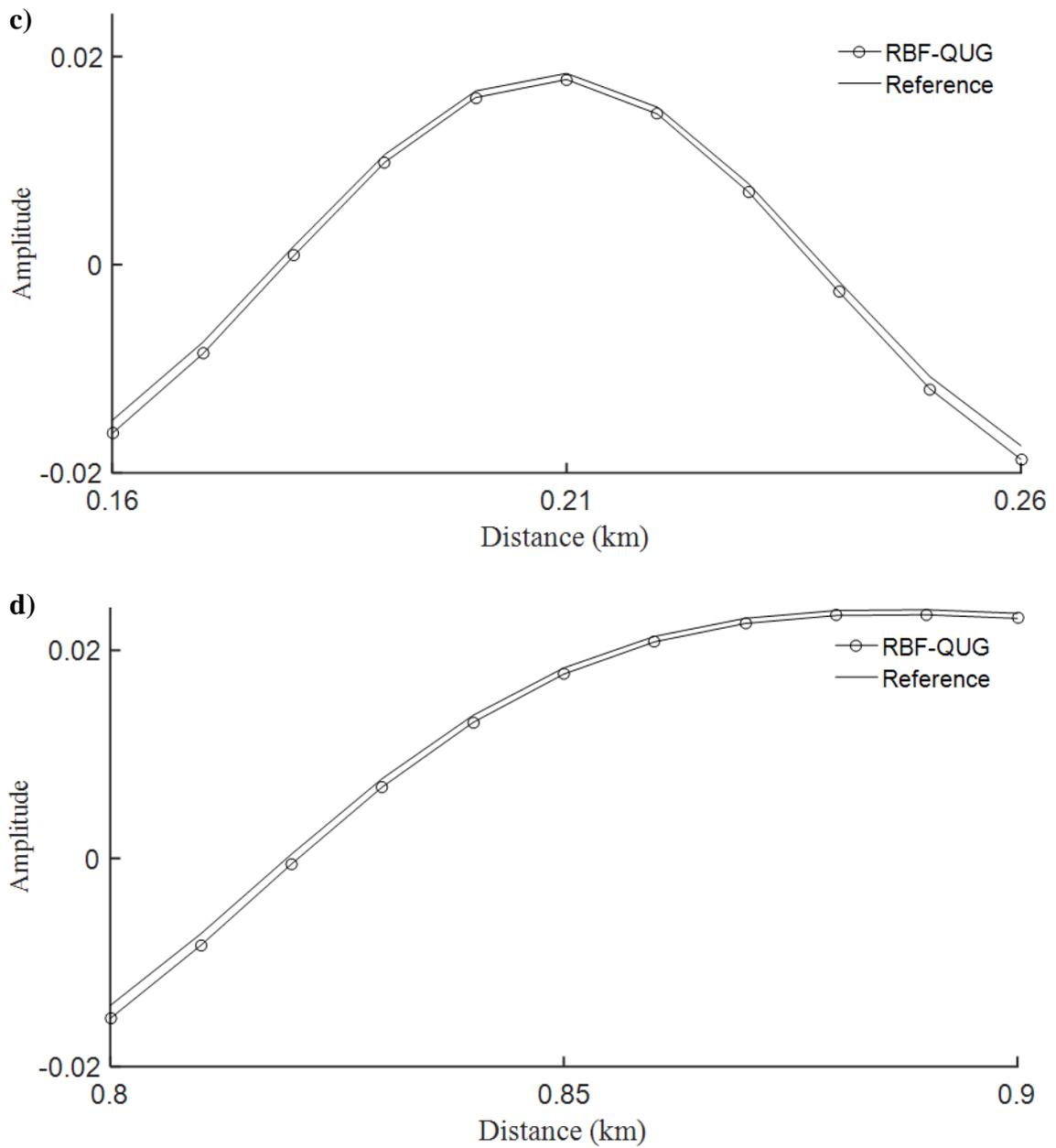

**Fig. 6.** The zoomed-in version of waveform Numerical solutions for three different node configuration schemes.

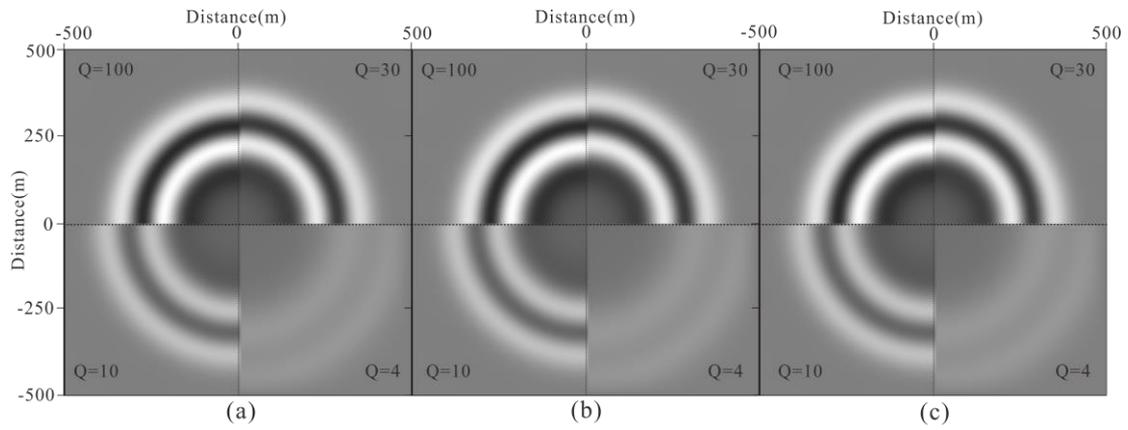

**Fig. 7.** Four wavefield parts corresponding to four $Q$ values: 100, 30, 10 and 4. The three subfigures from left to right computed by (a) pseudo-spectral method, (b) RBF collocation method based on the uniform node distribution and (c) RBF collocation method based on the quasi-uniform node distribution, respectively. The snapshots are recorded at 200ms in homogeneous attenuating media.

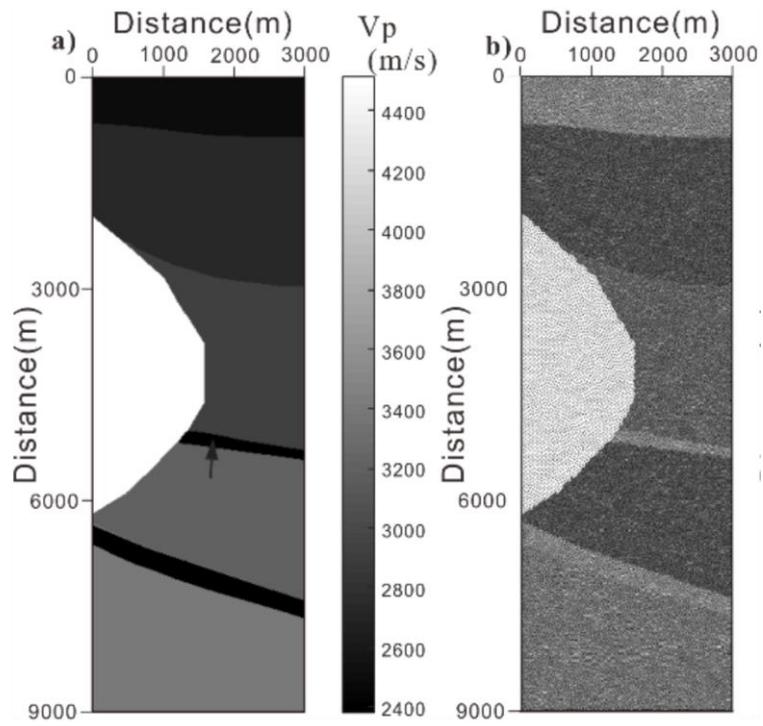
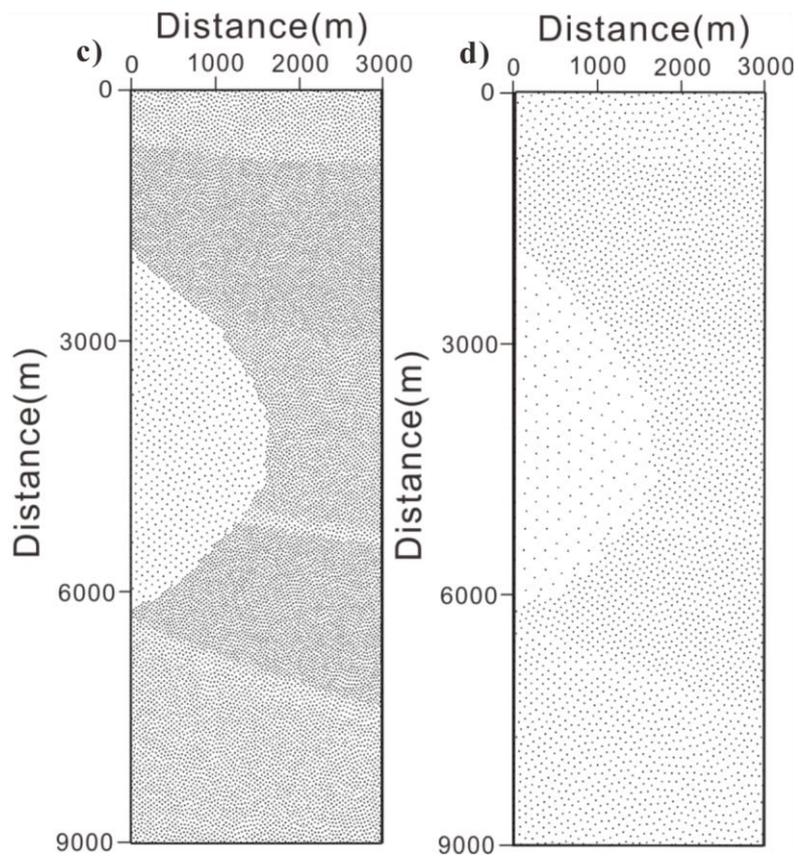

**Fig. 8.** Complex model (a) and its quasi-uniform node configuration. (b), (c), (d) are discretizations with increasing radii. The total numbers of nodes are 117412, 29494 and 13051 respectively.

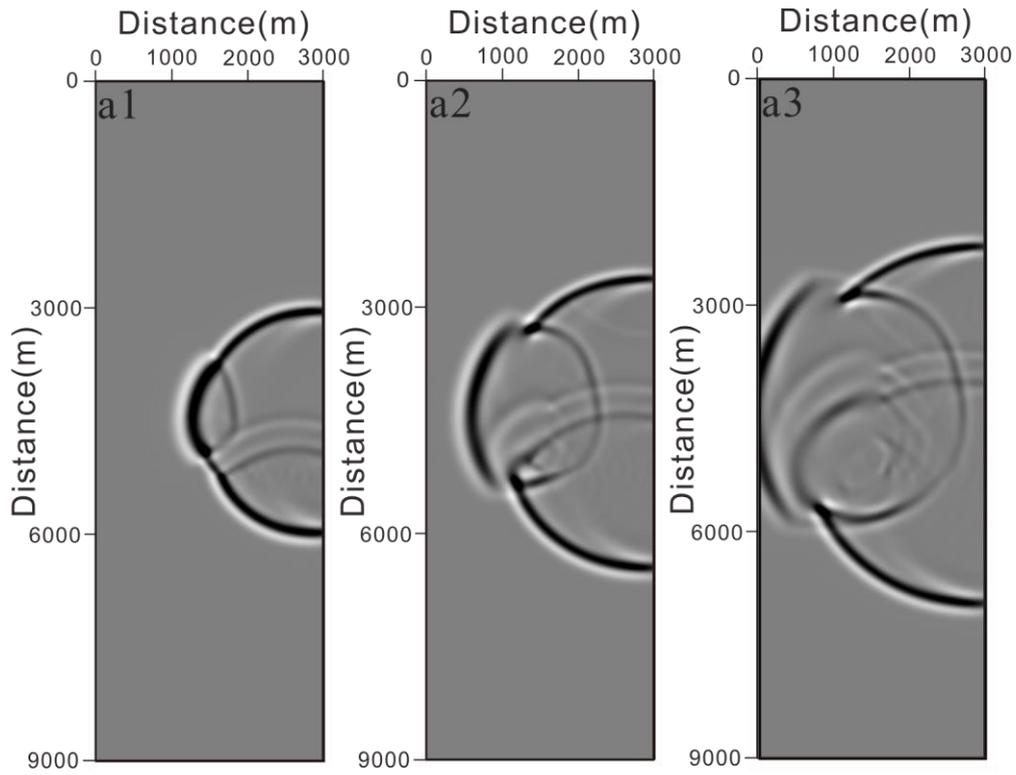
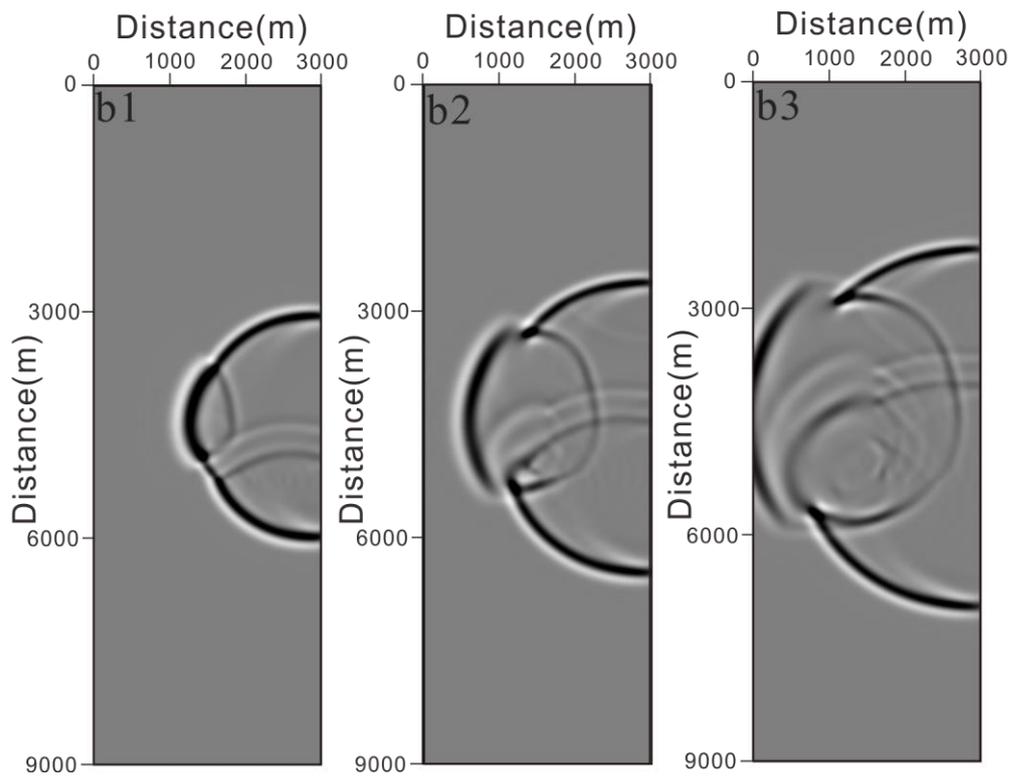

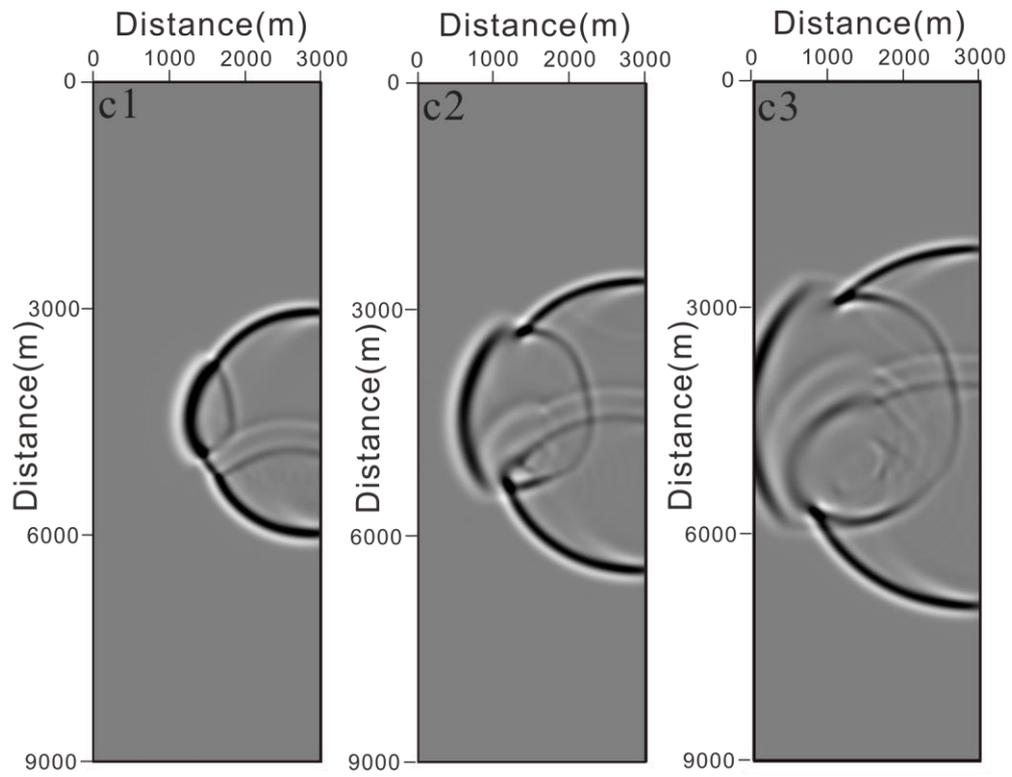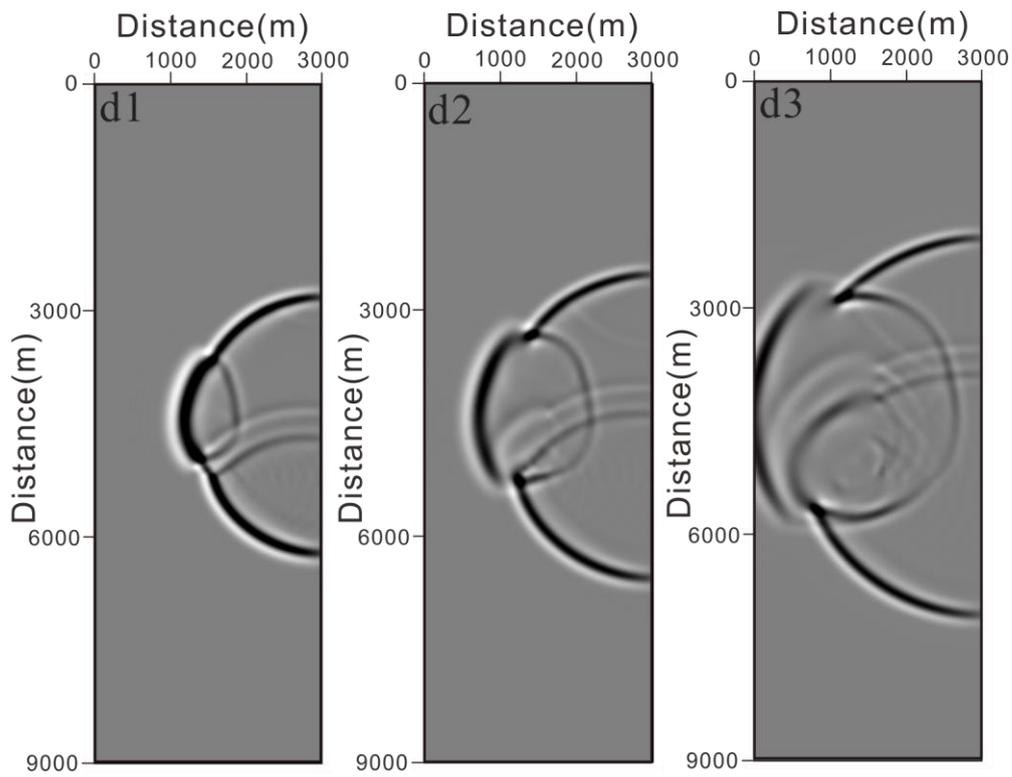

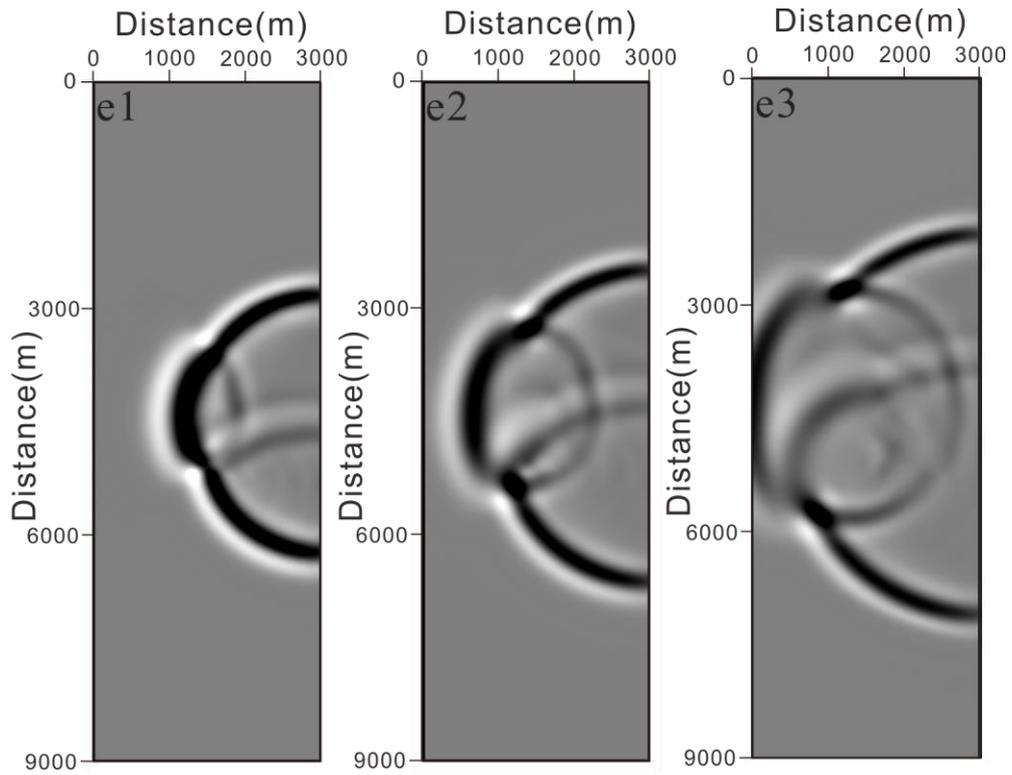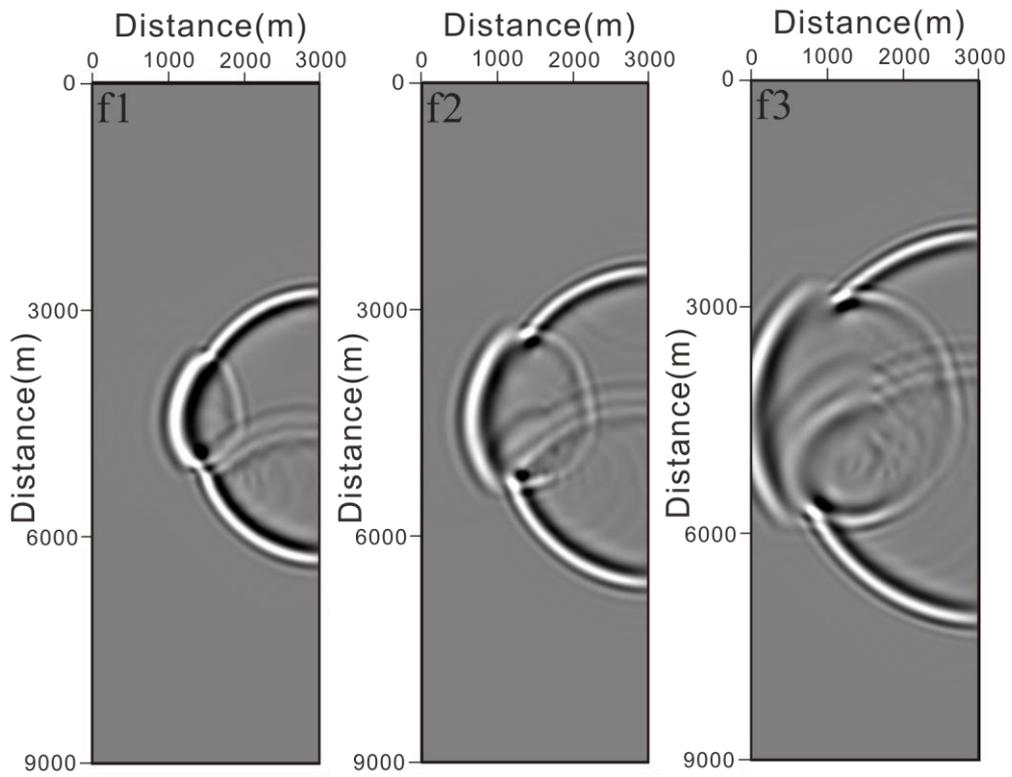

**Fig. 9.** The 2D partial SEG Hess model snapshots at $400ms$ (a1-f1), $600ms$ (a2-f2), $800ms$ (a3-f3). (a) The reference snapshots. (b) Snapshots computed by RBF collocation method with regular square uniform grids layout where $dx = dz = 7.5m$. (c) Snapshots computed by RBF collocation method with quasi-uniform grids layout where $r = r_{bs} = 5 \sim 10m$. (d) Snapshots computed by RBF collocation method with quasi-uniform grids layout where $r = 2r_{bs}$. (e) Snapshots computed by RBF collocation method with quasi-uniform grids layout where $r = 3r_{bs}$. (f) Snapshots computed by pseudo-spectral method with regular square uniform grids layout where $dx = dz = 30m$.

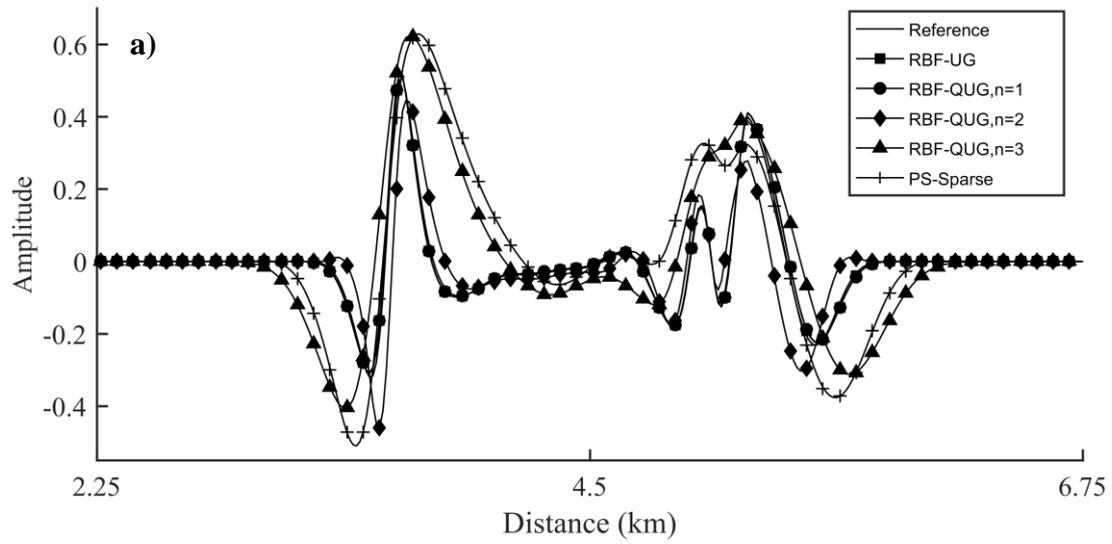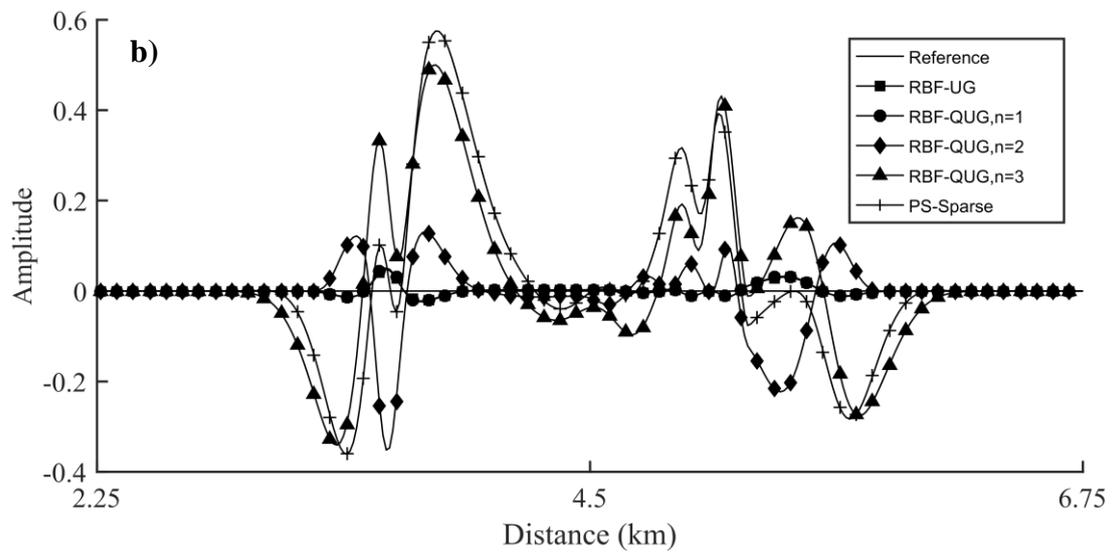

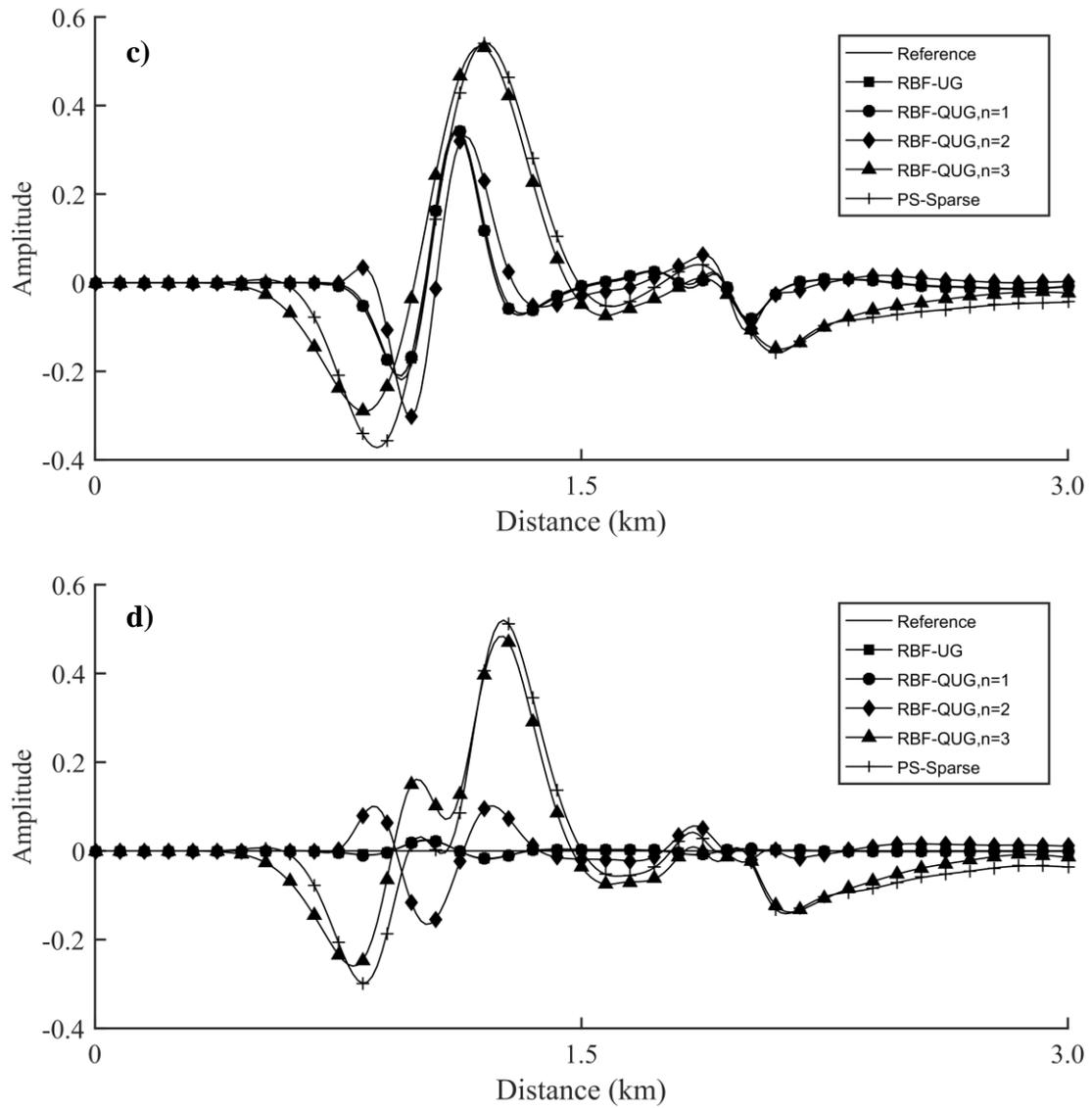

**Fig. 10.** Waveforms comparison among differenct methods for 2D partial SEG Hess model. (a) and (c) display the waveforms extracted from Fig. 9 at $400ms$, (c) and (d) show the difference of the various methods with reference result.

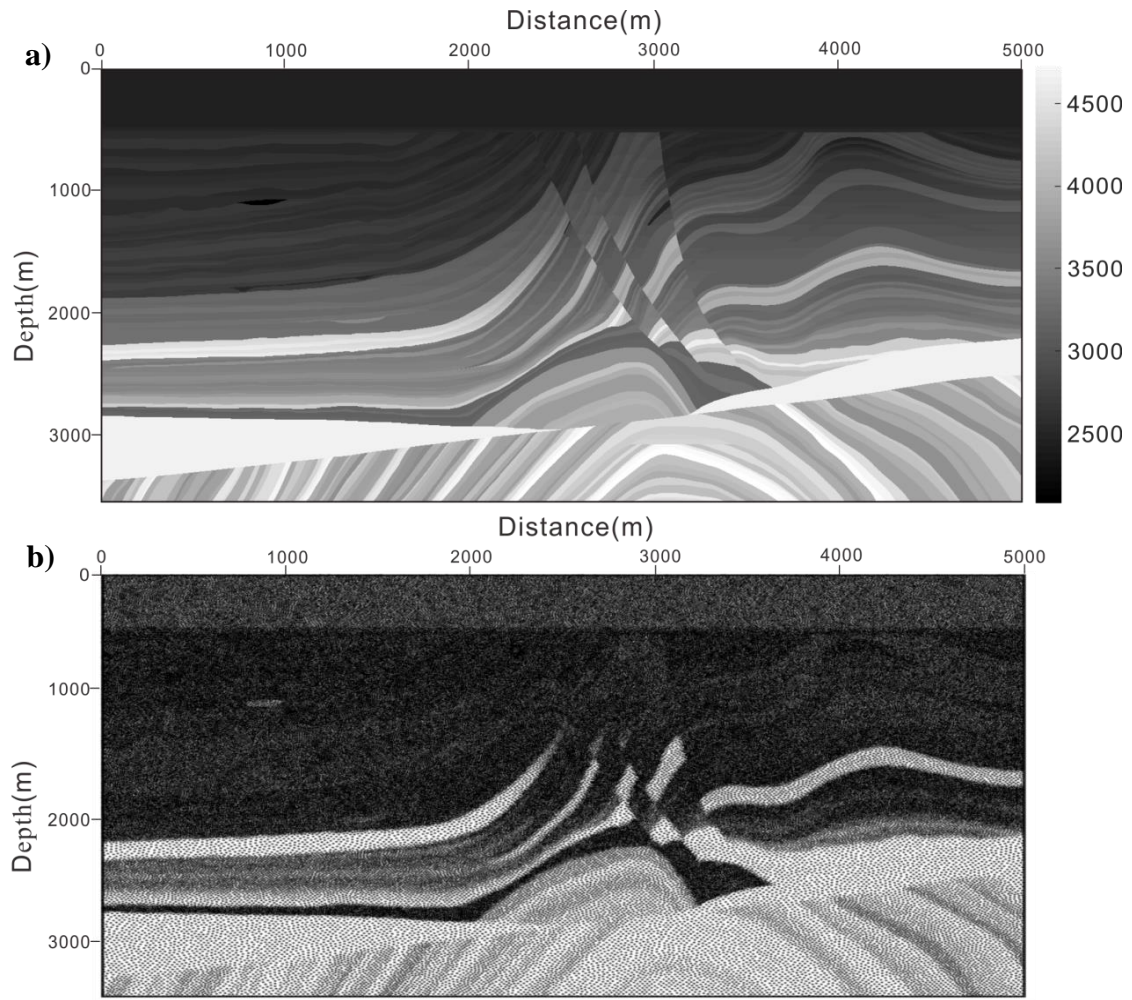

**Fig. 11.** The 2D Marmousi model and its quasi-uniform mesh-free nodal distribution. (a) The velocity model. (b) The quasi-uniform nodal distribution obtained by the fast-generation algorithm.[22]

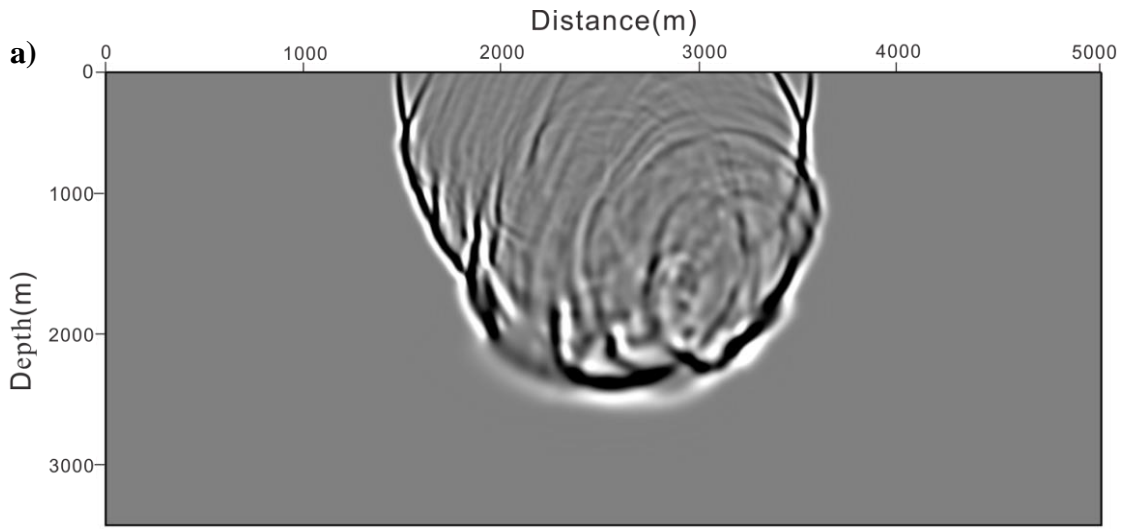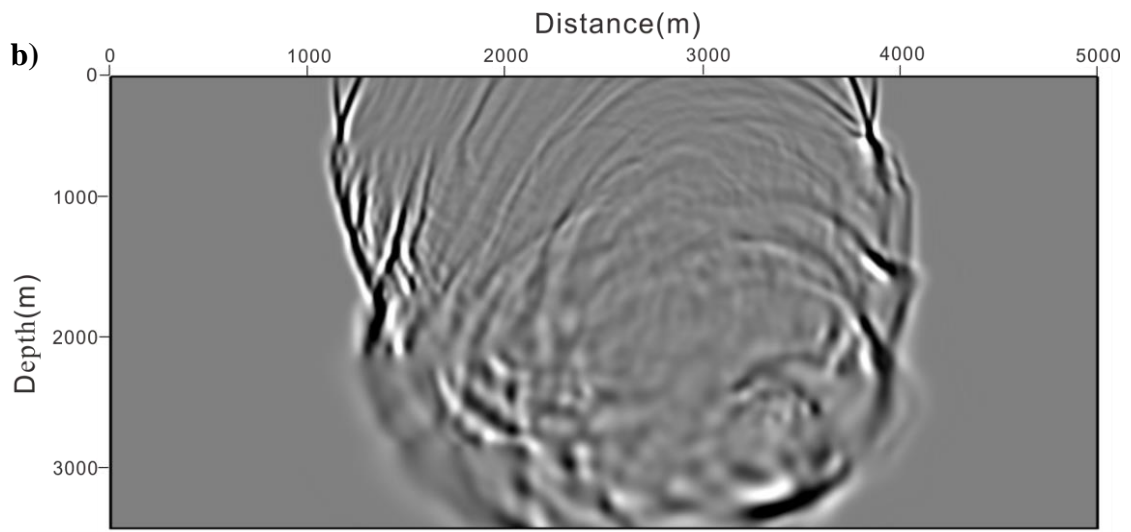

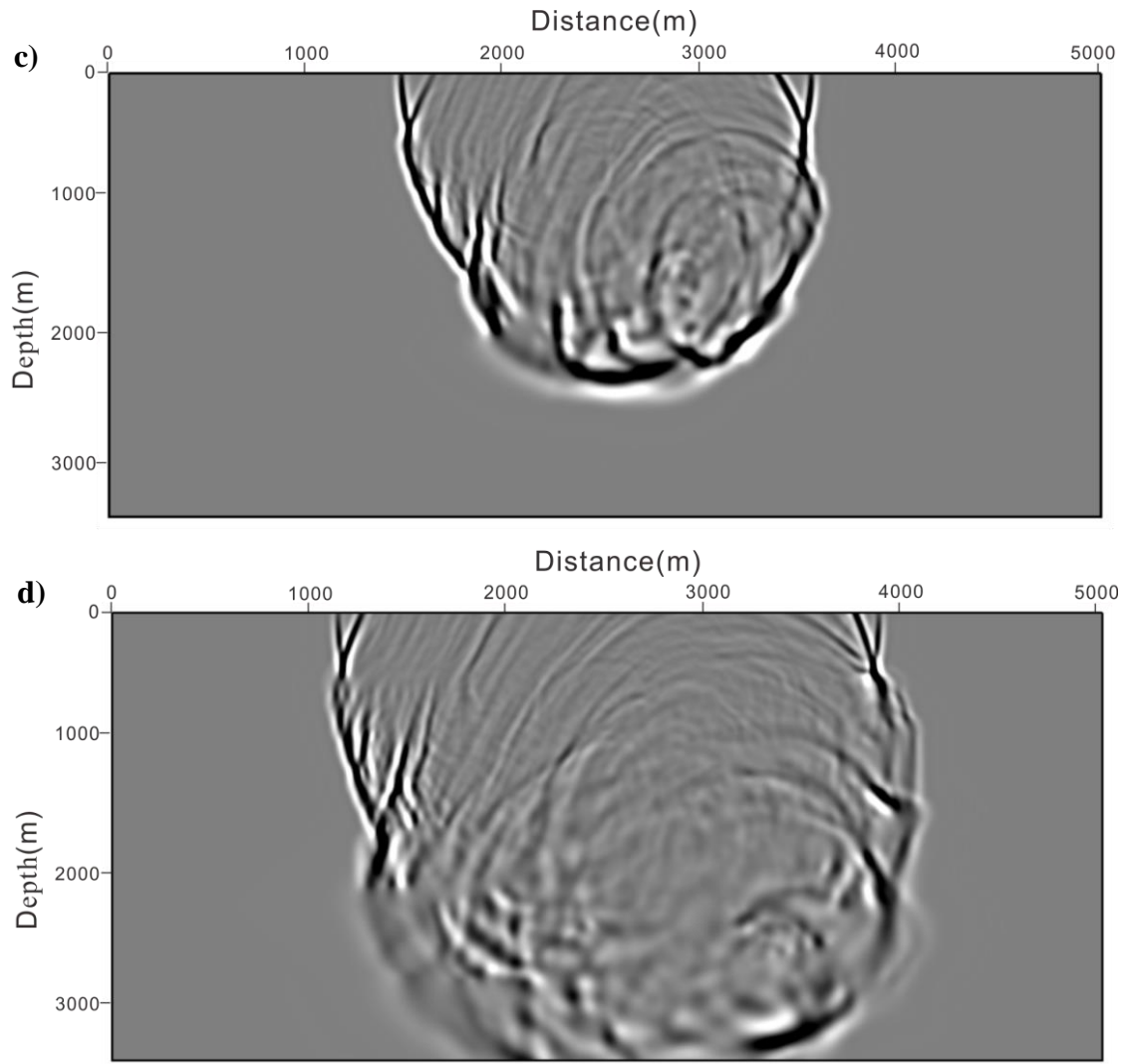

**Fig. 12.** The 2D modified Marmousi snapshots at $1s$ and $2s$, (a) and (b) computed by RBF collocation method with quasi-uniform nodes distribution where $r = 10 \sim 20m$ while (c) and (d) computed by pseudo-spectral method with regular square uniform nodes distribution where $dx = dz = 7.5m$.

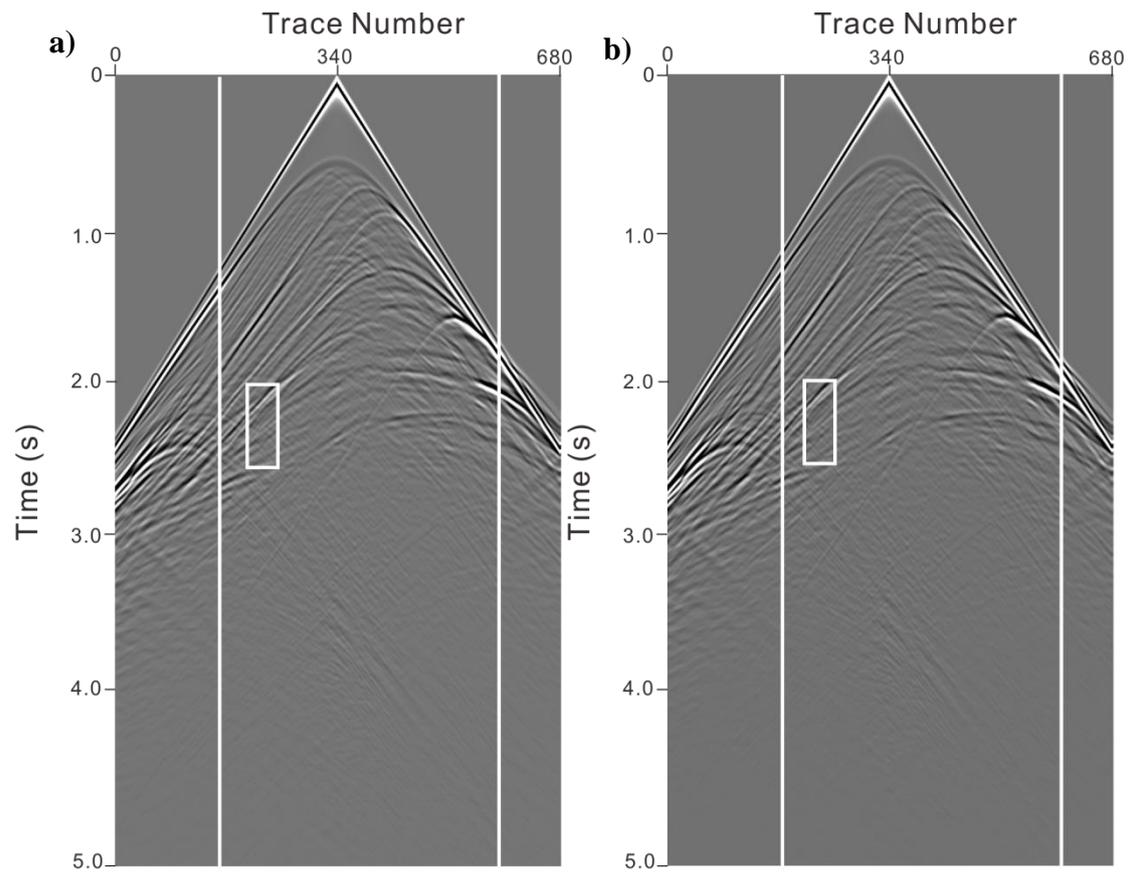
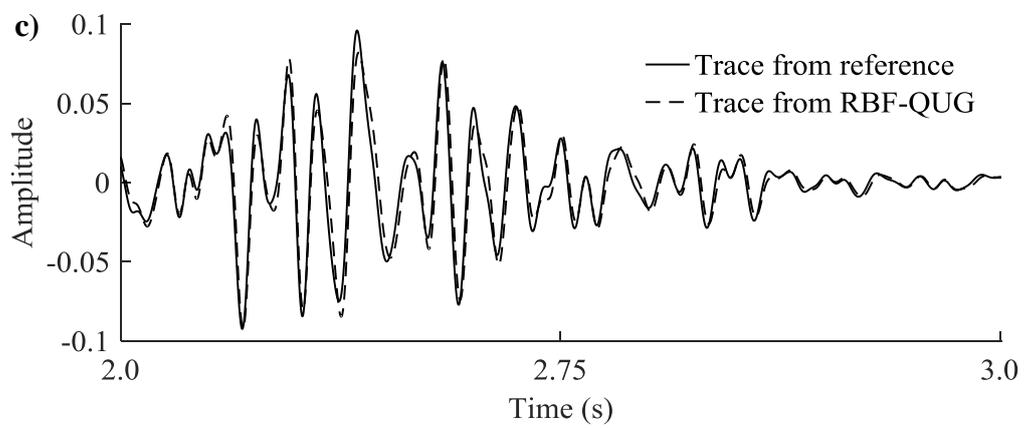
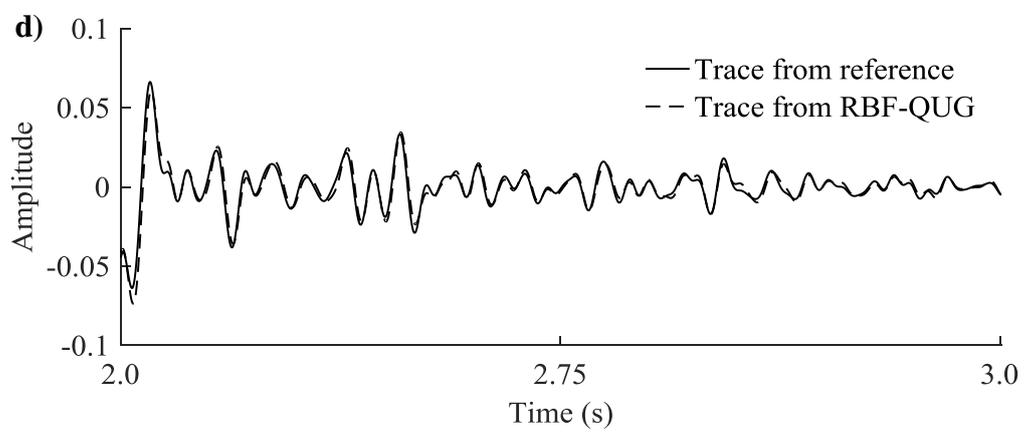

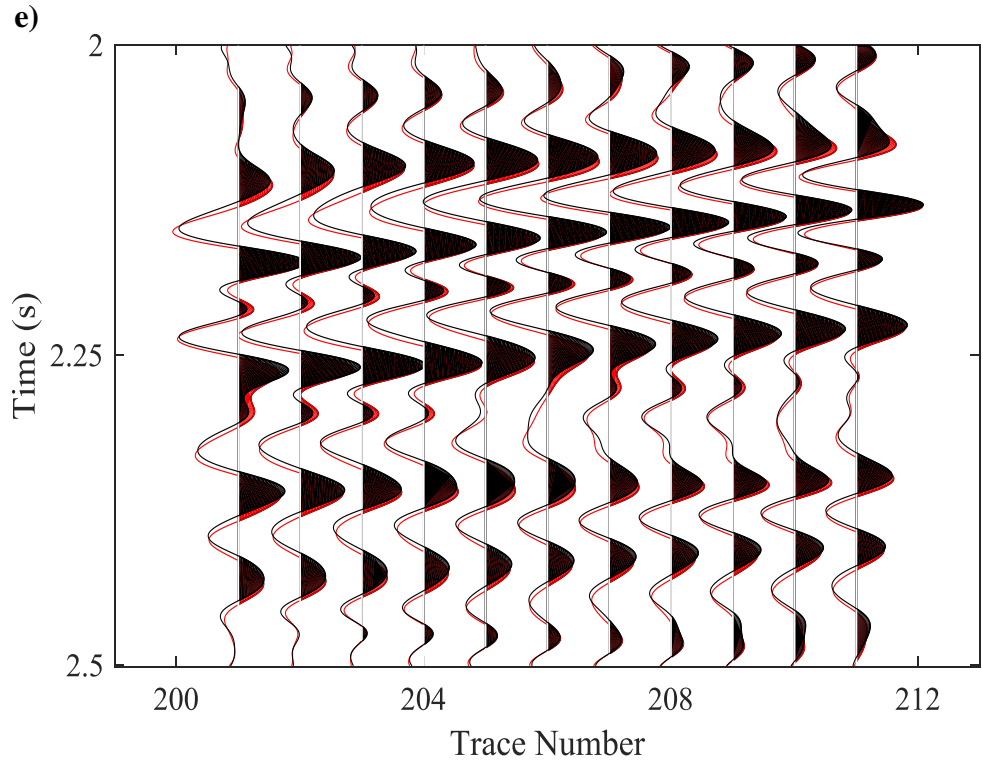

**Fig. 13.** (a-b) Seismograms from Fig. 10. (a) RBF method based on the quasi-uniform distribution of nodes and (b) pseudo-spectral method based on the uniform distribution of nodes. The subfigures (c) and (d) indicate the waveform comparison among different methods extracted from the (a) and (b) (white line indication). (c) Extracted from the left line and (d) extracted from the right line. (e) Local magnification of the white box in the seismograms (a,b) to highlight the local refraction waveforms. The red lines indicate the RBF method and the black lines indicate the pseudo-spectral method.

Table 1.

Commonly used types of radial basis functions (writing $r = \|\mathbf{x} - \mathbf{x}_i\|$).

| Types | RBF |
|---|---|
| Gaussian | $\phi(r) = e^{-(\varepsilon r)^2}$ |
| Multiquadric | $\phi(r) = \sqrt{1 + (\varepsilon r)^2}$ |
| Inverse quadratic | $\phi(r) = \dfrac{1}{1 + (\varepsilon r)^2}$ |
| Inverse multiquadric | $\phi(r) = \dfrac{1}{\sqrt{1 + (\varepsilon r)^2}}$ |
| Polyharmonic spline | $\phi(r) = r^k,\ k = 1, 3, 5, \ldots$ <br> $\phi(r) = r^k \ln(r),\ k = 2, 4, 6, \ldots$ |
| Thin plate spline | $\phi(r) = r^2 \ln(r)$ |

Table 2.

Comparison of calculation behavior between two different node configuration schemes.

| | Quasi-uniformly configuration scheme | | | Uniformly configuration scheme | | |
|---|---|---|---|---|---|---|
| Domain Size (km) | Grids number | CPU time (ms) | Relative error ($\times 10^{-3}$) | Grids number | CPU time (ms) | Relative error ($\times 10^{-3}$) |
| 0.5×0.5 | 2080 | 6249.112 | 3.586 | 2601 | 6692.331 | 3.692 |
| 1.5×1.5 | 13527 | 13862.265 | 0.652 | 22801 | 20043.521 | 0.691 |
| 2.5×2.5 | 30921 | 19921.591 | 0.429 | 63001 | 33691.207 | 0.477 |
| 5.0×5.0 | 11482 | 32547.911 | 0.304 | 251001 | 63132.113 | 0.389 |